\makeatletter
\newcommand{\linebreakand}{%
  \end{@IEEEauthorhalign} \hfill\mbox{}\par
  \mbox{}\hfill\begin{@IEEEauthorhalign}
}
\makeatother

\documentclass[conference]{IEEEtran}
\IEEEoverridecommandlockouts

\usepackage{fancyhdr}          % right after \documentclass[conference]{IEEEtran}

\usepackage{amsmath,amssymb,amsfonts}
\usepackage{amsthm}

\usepackage{algorithmic}
\usepackage{graphicx}
\usepackage[caption=false,font=footnotesize]{subfig}
\usepackage{textcomp}
\usepackage{xcolor}
\usepackage{colortbl}
\usepackage{hyperref}
\usepackage{cleveref}
\usepackage{tabularx}
\usepackage{pifont}
\usepackage{dblfloatfix}
\usepackage[inline]{enumitem}
\usepackage{tikz}
\usepackage{doi}

\usepackage[defaultlines=2,all]{nowidow}
\usepackage{multirow}
\usepackage{mdframed}
\usepackage{booktabs} % professional-quality tables (\toprule/\midrule/\bottomrule)
\usepackage{makecell} % compact multi-line table cells (\makecell)
\usepackage{threeparttable} % table notes directly under the table
\usepackage{siunitx}
\usepackage[backend=biber,style=ieee,doi=true,url=true]{biblatex}
\AtEveryBibitem{%
  \iffieldundef{doi}{}{%          % DOI present -> drop URL, arXiv link, and access date
    \clearfield{url}%
    \clearfield{urlyear}%
    \clearfield{urlmonth}%
    \clearfield{urlday}%
    \clearfield{eprint}%
    \clearfield{eprinttype}%
    \clearfield{eprintclass}%
  }%
}
\AtEveryBibitem{\clearfield{note}}

\mdfdefinestyle{calloutstyle}{%
  linewidth=0.6pt,
  roundcorner=2pt,
  innerleftmargin=6pt,
  innerrightmargin=6pt,
  innertopmargin=6pt,
  innerbottommargin=6pt,
  skipabove=6pt,
  skipbelow=6pt,
  backgroundcolor=black!3
}

\newenvironment{callout}[1][Takeaway]{%
  \begin{mdframed}[style=calloutstyle]%
  \noindent\textbf{#1: }\ignorespaces
}{%
  \unskip\end{mdframed}%
}

\DeclareUnicodeCharacter{03BB}{\ensuremath{\lambda}}

\def\BibTeX{{\rm B\kern-.05em{\sc i\kern-.025em b}\kern-.08em
    T\kern-.1667em\lower.7ex\hbox{E}\kern-.125emX}}
\begin{document}

\title{
What Bloats Your Floats? Right-Sizing Numerical Precision for Scientific Computing
}

\author{%
\IEEEauthorblockN{Pratyai Mazumder}
\IEEEauthorblockA{\textit{ETH Zurich} \\
Zurich, Switzerland \\
\mbox{pmazumder@ethz.ch}}
\and
\IEEEauthorblockN{Alexandru Calotoiu}
\IEEEauthorblockA{\textit{ETH Zurich} \\
Zurich, Switzerland \\
\mbox{acalotoiu@ethz.ch}}
\and
\IEEEauthorblockN{Marcin Copik}
\IEEEauthorblockA{\textit{ETH Zurich} \\
Zurich, Switzerland \\
\mbox{mcopik@ethz.ch}}
\linebreakand
\IEEEauthorblockN{Yakup Koray Budanaz}
\IEEEauthorblockA{\textit{ETH Zurich} \\
Zurich, Switzerland \\
\mbox{ybudanaz@ethz.ch}}
\and
\IEEEauthorblockN{Shriram Chandran}
\IEEEauthorblockA{\textit{ETH Zurich} \\
Zurich, Switzerland \\
\mbox{schandran@ethz.ch}}
\and
\IEEEauthorblockN{Torsten Hoefler}
\IEEEauthorblockA{\textit{ETH Zurich} \\
Zurich, Switzerland \\
\mbox{htor@ethz.ch}}

}

\maketitle
\thispagestyle{fancy}
\lhead{}\rhead{}\chead{}
\lfoot{\parbox{\textwidth}{\footnotesize \copyright~2026 IEEE. Personal use of this material is permitted. Permission from IEEE must be obtained for all other uses, in any current or future media, including reprinting/republishing this material for advertising or promotional purposes, creating new collective works, for resale or redistribution to servers or lists, or reuse of any copyrighted component of this work in other works.}}
\rfoot{}\cfoot{}
\renewcommand{\headrulewidth}{0pt}
\renewcommand{\footrulewidth}{0pt}

\begin{abstract}
While FP64 (\emph{binary64}) remains the standard representation of real numbers in scientific computing, hardware trends increasingly favor low-precision formats optimized for AI workloads. This shift prioritizes throughput and energy efficiency over full-precision capabilities. We challenge the necessity of high-precision arithmetic in PDE-governed simulations, where spatial and temporal discretization errors introduce a physical noise floor that masks the lower-order bits of the FP64 format, rendering them irrelevant. We propose a criterion based on the Signal-to-Noise Ratio (SNR) with respect to spatial and temporal refinements to evaluate the impact of safe precision reduction on scientific computations. Finally, we present an automated dataflow-centric workflow for targeted lowering within mixed-precision subgraphs in complex scientific applications. Large-scale evaluations of weather and climate applications show that strategic precision reduction on GPUs achieves up to $1.8\times$ speedup without compromising physical fidelity.
\end{abstract}


\section{Introduction}\label{sec:introduction}

The FP64 format has long been the default for large-scale scientific applications~\cite{bhola_deterministic_2024, higham_mixed_2022}, historically deemed necessary for computational stability~\cite{higham_accuracy_2002, trefethen_numerical_1997}. Modern hardware, however, is moving in the opposite direction: driven by AI workloads, GPUs increasingly optimize for FP32, FP16, BF16, and FP8, relegating native FP64 to a marginal fraction of peak throughput~\cite{dongarra_hardware_2024, nvidia_nvidia_2024}. For memory-bandwidth-bound PDE solvers---the computational core of weather and climate models---remaining locked to 64-bit data structures exacerbates the memory wall and bottlenecks throughput~\cite{walden_mixed_2019, unat_persistent_2025, fuhrer_near-global_2018}. Because data movement dominates execution time~\cite{fuhrer_near-global_2018, shalf_future_2020}, halving the bit-width effectively doubles memory bandwidth~\cite{karp_effects_2025, klower_low-precision_2021}, providing a direct path to higher-resolution or larger-ensemble simulations~\cite{karp_effects_2025, palmer_more_2014, vana_single_2017, klower_low-precision_2021}.

From a physical standpoint, the case for FP64 is weaker than commonly assumed. At practical grid resolutions, spatial and temporal discretization errors dominate the solution error budget, masking floating-point round-off and rendering the lower-order significand bits of FP64 irrelevant~\cite{palmer_more_2014, thornes_power_2018, weens_modeling_2022, klower_low-precision_2021, palmer_towards_2012, bauer_multigrid_2025}. Yet production codes remain conservatively locked to FP64~\cite{demeure_algorithm_2023, vanover_engineering_2025, hoerold_raptor_2025, bauer_quiet_2015, bauer_digital_2021}---not because the physics demands it, but because the community lacks scalable tools to validate that lower precision is safe~\cite{matsuoka_myths_2023}.

The validation bottleneck is twofold. First, large-scale simulation frameworks consist of millions of lines of legacy code with complex numerical dependencies; naively downcasting variables across regions with heterogeneous error sensitivities can introduce instability, catastrophic cancellation, or solver divergence~\cite{benz_dynamic_2012, kashi_mixed-precision_2024}. The combinatorial explosion of assigning precisions to thousands of variables renders manual optimization intractable~\cite{higham_mixed_2022, graillat_auto-tuning_2019}, and existing auto-tuning methods fail to scale to monolithic codebases~\cite{matsuoka_myths_2023}.

Second, existing precision-tuning frameworks rely on the wrong evaluation methodology. They bound pointwise errors against a high-precision reference trajectory~\cite{constantinides_automated_2025, demeure_algorithm_2023, rubio-gonzalez_precimonious_2013}, but in PDE solvers this reference shares the same discretization error as the test run. The resulting metrics measure the arithmetic gap between formats---not the physical accuracy of either solution. In chaotic systems, acceptable grid-scale noise causes these pointwise metrics to saturate rapidly~\cite{gmd-6-735-2013, zangl_icon_2015, jablonowski_baroclinic_2006}, rendering tuning constraints overly pessimistic and sometimes misleading~\cite{course_state_2023, klower_low-precision_2021, saffin_reducedprecision_2020}.
 
Compounding the tooling gap is a measurement gap. Validating precision reductions requires collapsing multidimensional error fields into a robust scalar metric, yet the standard choices are ill-suited to chaotic PDE solvers. Pointwise $L_p$ norms grow exponentially under chaotic trajectory divergence~\cite{gmd-6-735-2013, zangl_icon_2015, jablonowski_baroclinic_2006}; relative errors hit zero-crossing singularities~\cite{darulova_daisy_2018}; and distributional metrics such as Wasserstein distance or CRPS~\cite{karp_effects_2025, klower_low-precision_2021, vana_single_2017}, while physically meaningful, are too expensive to embed in an automated tuning loop. What is needed is a metric that is scale-free, tolerant of nonlinear dynamics, and computable from a single execution---without requiring a reference beyond FP64.

\subsection{Contributions}\label{subsec:contributions}
 
\subsubsection*{\textbf{C1. A New Perspective on Precision Tuning in Scientific Computing}}
We propose a variance-based Signal-to-Noise Ratio (SNR) criterion that replaces the conventional high-precision reference with the discretization noise floor as the evaluation baseline. This threshold is derived from spatial and temporal refinements alone, reflects the intrinsic accuracy limit of the discretization, and satisfies the requirements above: it is scale-free, tolerant of nonlinear dynamics, and needs no reference beyond FP64.
 
\subsubsection*{\textbf{C2. Automated Dataflow-Level Precision Tuning in Production-Scale Weather and Climate Models}}
We build an automated framework that performs systematic precision reduction directly on the dataflow graph and validates safety against the proposed threshold. We demonstrate its efficacy on the ECMWF CLOUDSC microphysics scheme~\cite{ecmwf_ecmwf-ifsdwarf-p-cloudsc_2025} and the ICON velocity tendency kernel~\cite{zangl_icon_2015}, achieving up to $1.8\times$ GPU speedup while preserving macroscopic numerical fidelity.

\section{Discretization Noise and Precision Limits}\label{sec:discretization-noise-and-precision-limits}

\begin{figure}[t]
    \centering
    \includegraphics[width=\linewidth]{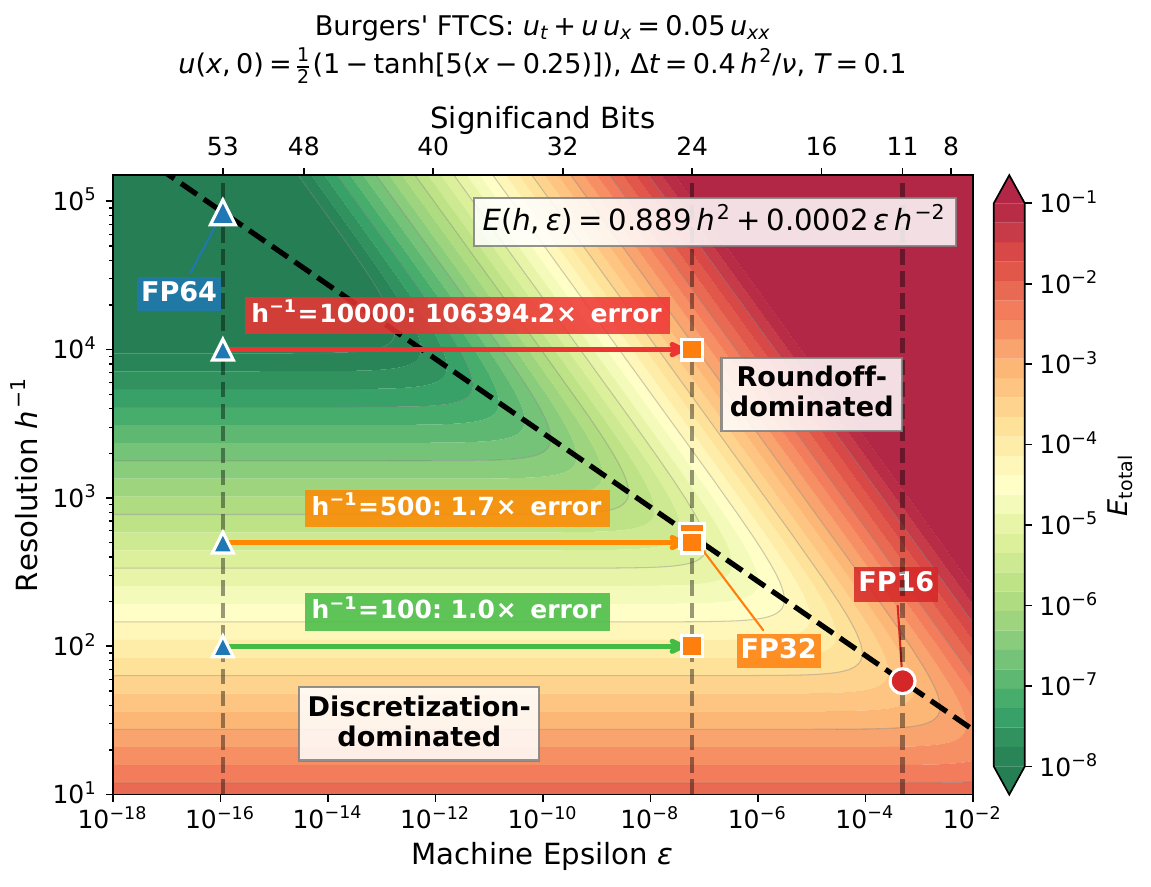}
    \caption{Total error landscape $E_{\text{total}} = \Theta(h^2) + \Theta(\epsilon\,h^{-2})$ for forward-time centered-space (FTCS) scheme on the 1D viscous Burgers' equation ($\nu=0.05$, $T=0.1$), demonstrating that the impact of lowering precision depends strongly on the operating grid resolution $h^{-1}$. The dashed line marks the optimal resolution locus $h^{-1}_{\text{opt}}(\epsilon)$. Arrows indicate the error penalty when switching from FP64 to FP32 at three distinct grid resolutions: the error increase is negligible at a coarse resolution of $h^{-1}=100$ (green), grows to $1.7\times$ at $h^{-1}=500$ (orange), and suffers a catastrophic ${\sim}10^5\times$ increase at a fine resolution of $h^{-1}=10\,000$ (red).}
    \label{fig:error_landscape_burgers}
\end{figure}

In scientific computing, physical phenomena are modeled by PDEs, and solved using discretization-based techniques like finite difference, finite volume, and finite element methods. While our work is motivated by weather and climate codes, the proposed techniques apply broadly across numerical solvers.

The total error of a PDE solver is conventionally bounded by the sum of the norms of the modeling error $E_{\text{model}}$, the discretization error $E_{\text{disc}}$, and the round-off error $E_{\text{ro}}$:
\begin{equation}\label{eq:error_decomposition}
    \|u^* - \hat{u}_{h, \tau}\| \le \|E_{\text{model}}\| + \|E_{\text{disc}}\| + \|E_{\text{ro}}\|,
\end{equation}
where $u^*$ is the true physical state and $\hat{u}_{h, \tau}$ is the computed solution using floating-point arithmetic for a uniform grid resolution $h$ and timestep $\tau$.

For a given physical model, we ignore $E_{\text{model}}$ and are concerned with $E_{\text{total}} \equiv E_{\text{disc}} + E_{\text{ro}}$. As a result, the computational accuracy is governed by the balance between the spatial discretization error and the round-off error. Consider a PDE of order $m$, with a numerical scheme of spatial order $p$ and temporal order $q$. The spatial and temporal discretization errors are $\mathcal{O}(h^p)$ and $\mathcal{O}(\tau^q)$, respectively. Because deterministic round-off error accumulates at a rate proportional to the number of timesteps, the round-off error is $\mathcal{O}(\epsilon \tau^{-1})$ at every grid point, provided that $\tau^{-1}\epsilon \ll 1$~\cite[Ch.~3]{higham_accuracy_2002}. Using continuous-equivalent discrete norms, one can show this same asymptotic bound extends to the global round-off error~\cite[App.~A]{leveque_finite_2007}. Consequently,
\begin{equation}\label{eq:total_error_h}
    \|E_{\text{disc}}\| + \|E_{\text{ro}}\| \le \mathcal{O}(h^p) + \mathcal{O}(\tau^q) + \mathcal{O}\left(\frac{\epsilon}{\tau}\right).
\end{equation}
We make two standard assumptions: (1) the stability constraints hold $\tau = \mathcal{O}(h^m)$, and (2) the scheme is balanced such that spatial error asymptotically bounds or matches temporal error ($p \le qm$)~\cite{leveque_finite_2007}. Then, minimizing the total error yields the optimal configuration: $h_{\text{opt}} = \mathcal{O}(\epsilon^{1/(p+m)})$.

However, hardware constraints restrict extreme-scale simulations to a much coarser resolution ($h_{\text{prac}} \gg h_{\text{opt}}$)~\cite{tamstorf_discretization-error-accurate_2021}. For instance, reaching $h_{\text{opt}}$ for FP64 in hydrodynamic simulations requires problem sizes orders of magnitude beyond exascale capabilities (e.g., 286 billion grid cells for a 1D Sod's shock tube)~\cite{weens_modeling_2022}. We can define the precision $\epsilon^{\ast} = \mathcal{O}(h_{\text{prac}}^{p+m})$ for a given resolution $h_{\text{prac}}$ at which the round-off error matches the discretization error. Clearly, at a precision $\epsilon \le \epsilon^{\ast}$, we can control the total error asymptotically with only the discretization error: $\|E_{\text{disc}}\| \le \|E_{\text{total}}\| \le 2\|E_{\text{disc}}\|$.

\Cref{fig:error_landscape_burgers} illustrates this balance between $h$ and $\epsilon$ for the 1D viscous Burgers' equation. At relatively coarse resolutions (e.g., $h^{-1} \le 500$), the penalty for reducing precision from FP64 to FP32 is almost negligible. Only at extreme refinement (e.g., $h^{-1}=10\,000$), well past the optimal locus, does the penalty become catastrophic.

\subsection{The Marginal Utility of Precision}
\label{subsec:marginal-utility-of-precision}

\begin{table}[!t]
  \centering
  \begin{threeparttable}
  \caption{\normalfont\small%
    Comparison of scalar, field-level error metrics for evaluating
    reduced-precision arithmetic in nonlinear PDEs.
    Computational cost is $\mathcal{O}(N)$ in all cases except the
    distributional metrics, which incur higher computational cost.
    \checkmark~=~yes,
    ${\sim}$~=~partial,
    \ding{55}~=~no.\\
    Columns: \textbf{Scale-free}~=~Dimensionless; adapts to dynamic range without manual normalization or zero-crossing pathologies;
    \textbf{Nonlinear-safe}~=~Tolerates shocks and chaotic divergence without false saturation;
    \textbf{Mean-invariant}~=~Insensitive to static mean shifts; evaluates anomalies rather than absolute states.%
  }
  \label{tab:metric_comparison}
  \newcommand{\yes}{\textcolor{green!60!black}{\ding{51}}}
  \newcommand{\no}{\textcolor{red}{\ding{55}}}
  \newcommand{\ptl}{\textcolor{orange!80!black}{$\boldsymbol{\sim}$}}

  \renewcommand{\arraystretch}{1.15}
  \setlength{\tabcolsep}{1.5pt}
  {\footnotesize
  \begin{tabularx}{\columnwidth}{@{}
      >{\arraybackslash}p{0.45\columnwidth}
      >{\centering\arraybackslash}X
      >{\centering\arraybackslash}X
      >{\centering\arraybackslash}X
  @{}}
    \toprule
    \textbf{Metric}
      & \makecell{\textbf{Scale-}\\\textbf{free}}
      & \makecell{\textbf{Non-}\\\textbf{linear-}\\\textbf{safe}}
      & \makecell{\textbf{Mean-}\\\textbf{invariant}} \\
    \midrule
    \addlinespace[2pt]
    \midrule
    \multicolumn{4}{@{}c@{}}{\footnotesize\itshape Pointwise \& norm-based} \\
    \midrule
    \makecell[l]{Max relative error\\($\|e/u\|_{L_\infty}$; ULPs)}
      & \ptl~\cite{izycheva_sound_2017}
      & \no\tnote{1}
      & \no \\
    \makecell[l]{Absolute max error ($\|e\|_{L_\infty}$)}
      & \no
      & \no\tnote{1}
      & \no \\
    \makecell[l]{Global error norms\\($\|e\|_{L_1}$, $\|e\|_{L_2}$; RMSE)}
      & \no
      & \ptl\tnote{1}
      & \no \\
    \addlinespace[2pt]
    \midrule
    \multicolumn{4}{@{}c@{}}{\footnotesize\itshape Distributional} \\
    \midrule
    \makecell[l]{Wasserstein (optimal transport)}
      & \no
      & \yes
      & \no~\cite{peyre_computational_2020} \\
    \makecell[l]{CRPS (ensemble scoring)}
      & \no
      & \yes
      & \no \\
    \makecell[l]{Hellinger ($f$-divergence)}
      & \yes
      & \yes
      & \no \\
    \makecell[l]{KS distance (supremum CDF gap)}
      & \yes
      & \yes
      & \no \\
    \addlinespace[2pt]
    \midrule
    \multicolumn{4}{@{}c@{}}{\footnotesize\itshape Perceptual \& Structural} \\
    \midrule
    \makecell[l]{SSIM (image quality)}
      & \no~\cite{baker_dssim_2023}
      & \ptl
      & \no~\cite{wang_image_2004} \\
    \makecell[l]{DSSIM (data SSIM)}
      & \yes~\cite{baker_dssim_2023}
      & \ptl
      & \no~\cite{wang_image_2004} \\
    \addlinespace[2pt]
    \midrule
    \multicolumn{4}{@{}c@{}}{\footnotesize\itshape Signal-to-noise} \\
    \midrule
    $\text{SNR}_{\text{pow}}$
      & \yes
      & \ptl
      & \no \\
    \rowcolor{green!5}
    $\text{SNR}_{\text{var}}$
      & \yes
      & \yes
      & \yes \\
    \bottomrule
  \end{tabularx}

  \begin{tablenotes}[flushleft]
    \footnotesize
    \item[1] Pointwise metrics fail to converge at shocks $\mathcal{O}(1)$ \cite{leveque_finite_2007}, and rapidly saturate globally under chaotic trajectory divergence \cite{palmer_more_2014, jablonowski_baroclinic_2006}. ($L_1$ provides partial tolerance by integrating $|e|$ rather than $e^2$).
  \end{tablenotes}

  }
  \end{threeparttable}
\end{table}

\begin{figure}[!tb]
    \centering
    \subfloat[Spatial refinement at a proportional CFL ratio.]{%
        \includegraphics[width=\linewidth]{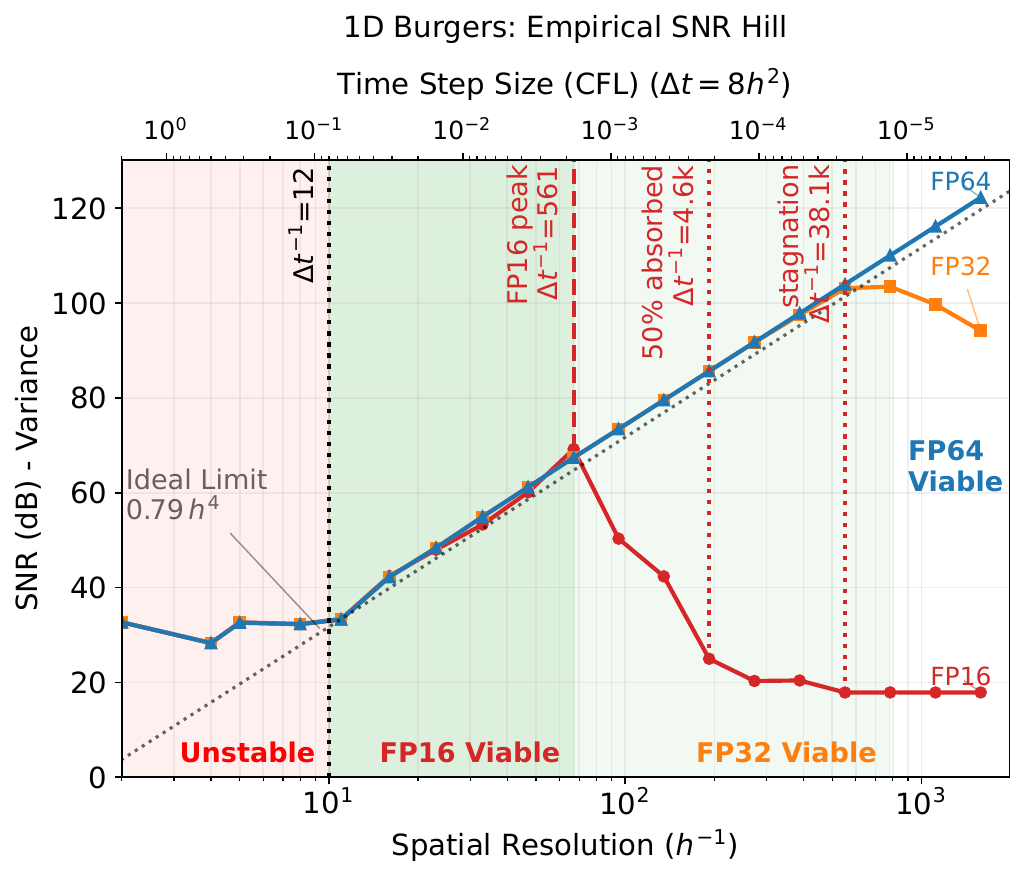}%
        \label{fig:ucurve-SNR-space}%
    }\hfill
    \subfloat[Temporal refinement at a fixed spatial resolution ($h=0.02$).]{%
        \includegraphics[width=\linewidth]{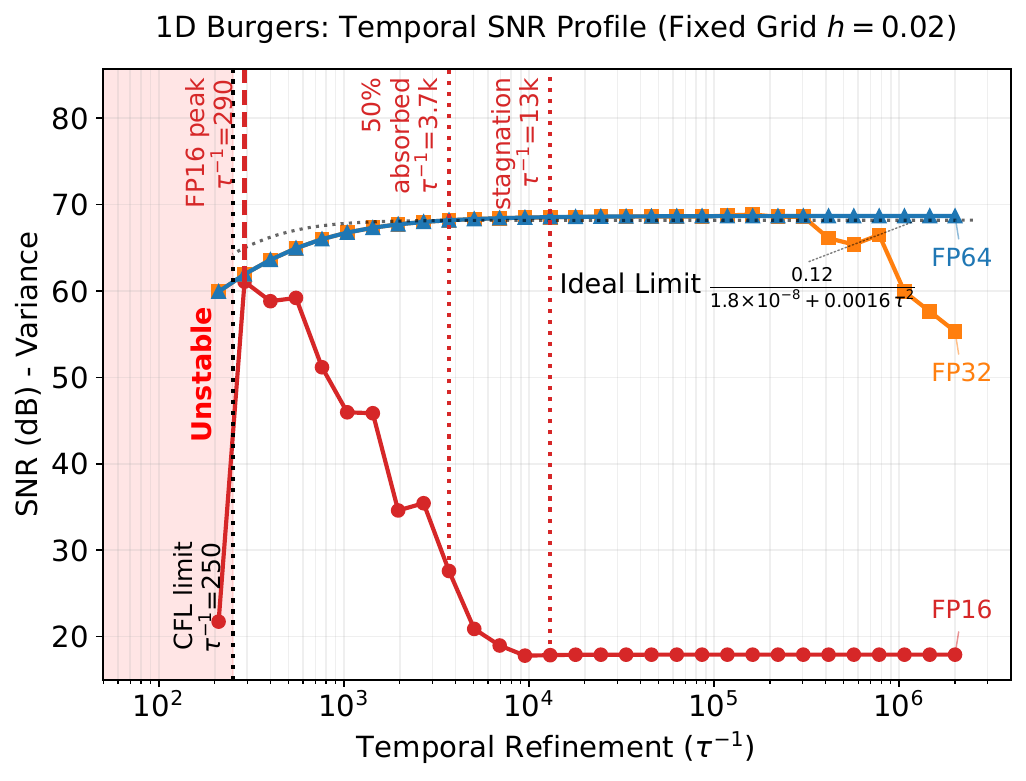}%
        \label{fig:ucurve-SNR-time}%
    }
        \caption{SNR curves for the 1D viscous Burgers' equation across three precisions. Lines show empirical SNR; the dotted \emph{Ideal Limit} is the precision-independent discretization floor. (a)~All precisions track the ideal $\mathcal{O}(h^{-4})$ slope until round-off ($\propto \varepsilon^2 h^{-4}$) dominates, producing a peaked hill that shifts rightward with precision. Past the peak, SNR does not collapse immediately: numerical viscosity partially absorbs round-off energy, creating a plateau between the ``50\% absorbed'' and ``stagnation'' markers before full degradation sets in. (b)~At fixed $h$, temporal refinement improves SNR as $\mathcal{O}(\tau^{-2})$ until saturating at the spatial discretization floor, then degrades through temporal round-off ($\propto \varepsilon^2 \tau^{-2}$), with an analogous absorption plateau for FP32.}
    \label{fig:ucurve-SNR}
\end{figure}

To understand the utility of precision beyond $\epsilon^\ast$, we model the round-off error as \emph{noise}. Although FP arithmetic is deterministic, the propagation of round-off errors in fluctuating dynamic signals is classically modeled via Pseudo-Quantization Noise (PQN) theory~\cite{widrow_quantization_2008, hull_tests_1966}. Under this PQN framework, the round-off error $E_{\text{ro}}(\epsilon)$ for a given precision $\epsilon$ behaves as a zero-mean stochastic variable that is uncorrelated with the discretization error $E_{\text{disc}}$.

The statistics of the total error $E_{\text{total}}$ reduce to a mean of $E_{\text{disc}}$ and a variance of $\mathbb{E}[\|E_{\text{ro}}(\epsilon)\|^2]$, and the second moment accumulates in quadrature:
\begin{equation}
    \mathbb{E}[\|E_{\text{total}}(\epsilon)\|^2] = \|E_{\text{disc}}\|^2 + \mathbb{E}[\|E_{\text{ro}}(\epsilon)\|^2].
\end{equation}
Furthermore, the round-off variance scales quadratically with machine epsilon: $\mathbb{E}[\|E_{\text{ro}}(\epsilon)\|^2] \propto \epsilon^2$~\cite[Ch.~12]{widrow_quantization_2008}.

Next, we evaluate the utility of any higher precision format $\epsilon_k = 2^{-k}\epsilon^{\ast}$ using $k$ additional significand bits. Recalling that $\mathbb{E}[\|E_{\text{ro}}(\epsilon^{\ast})\|^2] = \|E_{\text{disc}}\|^2$ as per the definition of $\epsilon^{\ast}$, the second moment of the total error becomes:
\begin{equation}
    \mathbb{E}[\|E_{\text{total}}(\epsilon_k)\|^2] = \|E_{\text{disc}}\|^2 \left(1 + 4^{-k}\right).
\end{equation}
The marginal variance reduction provided by the $k$-th additional bit after $\epsilon^{\ast}$ decays exponentially as $\mathcal{O}(4^{-k})$:
\begin{equation}
    \Delta_k = \mathbb{E}[\|E_{\text{total}}(\epsilon_{k-1})\|^2] - \mathbb{E}[\|E_{\text{total}}(\epsilon_k)\|^2] = 3 \cdot 4^{-k} \|E_{\text{disc}}\|^2.
\end{equation}
Consequently, the total error converges exponentially to the discretization floor $\|E_{\text{disc}}\|$. The additional significand bits required to reach $\epsilon_{64}$ from $\epsilon^{\ast}$ provide diminishing returns for the macroscopic solution fidelity.

\subsection{Physical Limits and Error Quantification}\label{subsec:physical-limits}

To quantify numerical fidelity, the multidimensional error field must be collapsed into a scalar metric. As summarized in \Cref{tab:metric_comparison}, traditional deterministic bounds fail to capture the physical stability of nonlinear PDEs. By treating round-off error $E_{\text{ro}}$ as \emph{noise}, we transition to the Signal-to-Noise Ratio (SNR). The conventional power-based SNR is:
\begin{equation}\label{eq:snr_pow}
    \text{SNR}_{\text{pow}}[u^*, \hat{u}_{h, \tau}] = \frac{\mathbb{E}[\|u^*\|^2]}{\mathbb{E}[\|u^* - \hat{u}_{h, \tau}\|^2]},
\end{equation}
where $u^*$ and $\hat{u}_{h, \tau}$ are the analytical and computed solutions, respectively. Under spatial averaging, $\text{SNR}_{\text{pow}}$ corresponds strictly to the inverse squared relative $L_2$ norm. While this expectation-based formulation effectively reframes geometric distance as stochastic noise power, both $\text{SNR}_{\text{pow}}$ and the $L_2$ norm remain highly sensitive to harmless static biases.

Because dynamic systems are driven by spatiotemporal gradients rather than absolute magnitudes, the variance-based SNR~\cite{kalliojarvi_roundoff_1996} discards the bias to promote the structural preservation of the dynamic anomalies and fluctuations that govern the simulated physics:
\begin{equation}\label{eq:snr_var}
    \text{SNR}_{\text{var}}[u^*, \hat{u}_{h, \tau}] = \frac{\text{Var}[u^*]}{\text{Var}[u^* - \hat{u}_{h, \tau}]}.
\end{equation}
For uncorrelated discretization and round-off errors, the denominator of \Cref{eq:snr_var} is decomposed as:
\begin{equation}
\text{Var}[E_{\text{total}}] \equiv \text{Var}[u^* - \hat{u}_{h, \tau}] = \text{Var}[E_{\text{disc}}(h)] + \text{Var}[E_{\text{ro}}(\epsilon)].
\end{equation}
For a spatial discretization of order $p$, $\text{Var}[E_{\text{disc}}(h)] = C_1 h^{2p}$. Under the PQN framework, the round-off variance scales as $\text{Var}[E_{\text{ro}}(\epsilon)] = C_2 \epsilon^2 h^{-2m}$. Substituting these into \Cref{eq:snr_var} yields the scaling relationship:
\begin{equation}\label{eq:snr_scaling}
\text{SNR}_{\text{var}}(\epsilon, h) = \frac{\text{Var}[u^*]}{C_1 h^{2p} + C_2 \epsilon^2 h^{-2m}},
\end{equation}
where $C_1$ and $C_2$ are problem-dependent proportionality constants, and often unknowable analytically. Asymptotically, \Cref{eq:snr_scaling} suggests the ``hill'' shape of SNR-vs.-$h$ plot in \Cref{fig:ucurve-SNR-space}, and maximizing the function yields an optimal $h$ that is the maximum refinement a given precision $\epsilon$ can support. In the limit of infinite precision, the round-off noise vanishes, and we recover a more standard scaling curve:
\begin{equation*}
\lim_{\epsilon \to 0} \text{SNR}_{\text{var}}(\epsilon, h) = \frac{\text{Var}[u^*]}{C_1 h^{2p}} = \Theta(h^{-2p}).
\end{equation*}

Even in a purely chaotic system subject only to perturbation error, decorrelated trajectories saturate at twice the natural variance~\cite{dalcher_error_1987, savijarvi_error_1995}, bounding $\text{SNR}_{\text{var}}$ within $[-3, \infty)$~dB. Thus, $\text{SNR}_{\text{var}}$ serves as an effective descriptor of the system's inherent ``noise'' in a scale-free manner.

\subsection{An Example: SNRs in 1D Burgers' Equation}\label{subsec:empirical-validation}

To validate the theoretical bounds, we consider the 1D viscous Burgers' equation, $u_t + u\,u_x = \nu\,u_{xx}$, a canonical nonlinear parabolic PDE ($d=1$, $m=2$) that couples advective steepening with viscous diffusion~\cite{bonkile_systematic_2018}. We discretize this PDE using a forward-time centered-space (FTCS) scheme ($p=2$, $q=1$), whose diffusive stability limit constrains $\tau = \mathcal{O}(h^2)$. This scheme is deliberately minimal: its central-difference advection term is unconditionally unstable and relies entirely on physical viscosity for damping~\cite[Ch.~10]{leveque_finite_2007}, making it acutely sensitive to grid-scale noise injected by finite-precision arithmetic.

\Cref{fig:ucurve-SNR} presents the resulting SNR curves. In \Cref{fig:ucurve-SNR-space}, the empirical SNR profiles ($\text{SNR} \propto E^{-2}$) all closely track the $\mathcal{O}(h^{-4})$ spatial discretization limit at practical resolutions ($h^{-1} < 50$), confirming that surplus precision offers no measurable benefit to solution fidelity. An SNR collapse only occurs during extreme over-refinement, where accumulated round-off error ($\propto \varepsilon^2 h^{-4}$) overtakes the discretization limit. The same holds in time (\Cref{fig:ucurve-SNR-time}): refinement improves SNR only until the spatial floor caps it.

\subsection{Moving Beyond Arithmetic Baselines}\label{subsec:moving-beyond-arithmetic-baselines}

\begin{figure}[!tb]
    \centering
    \includegraphics[width=\columnwidth]{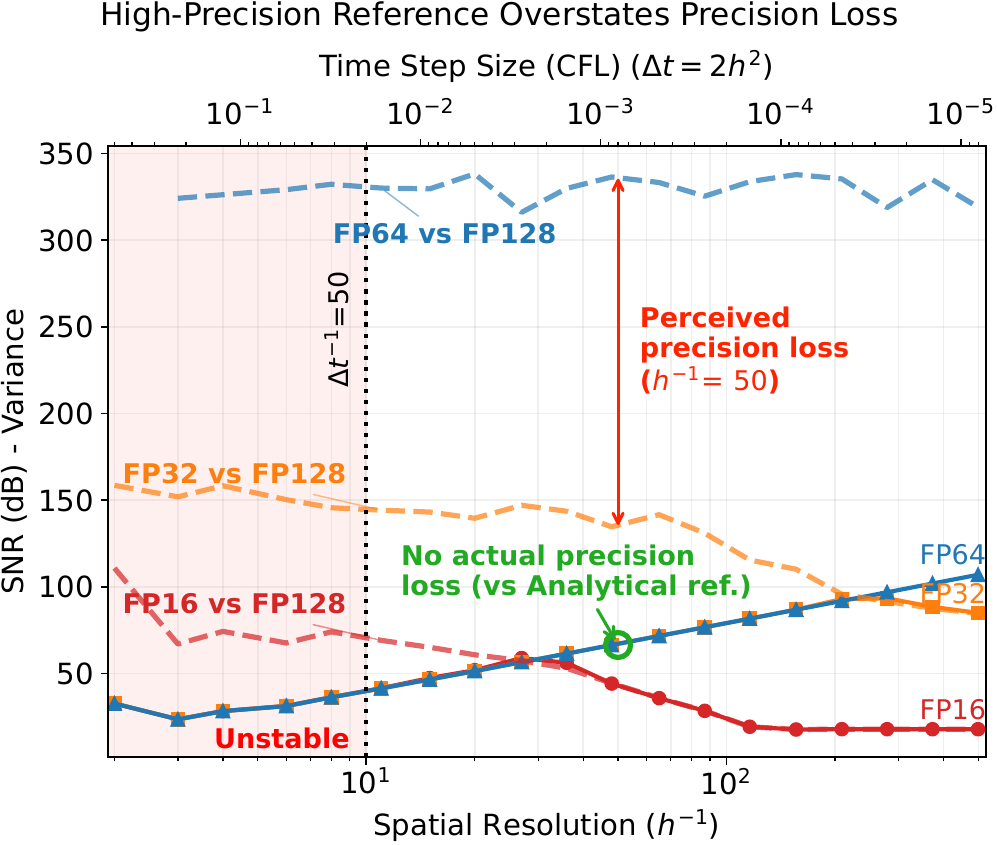}
    \caption{SNR of the 1-D Burgers FTCS solution measured against two baselines: the analytical solution (solid) and an FP128 numerical reference on the same grid (dashed). The red arrow marks the apparent FP32-to-FP64 gap in the reference curves at $h^{-1}=50$; the green circle marks the corresponding analytical SNR, where FP64 and FP32 are indistinguishable.}
    \label{fig:perception-snr}
    \vspace{-10pt}
\end{figure}

\begin{figure*}[!tbph]
    \centering
    \subfloat[Coarse grid ($N_x=128$)]{%
        \includegraphics[width=0.49\textwidth]{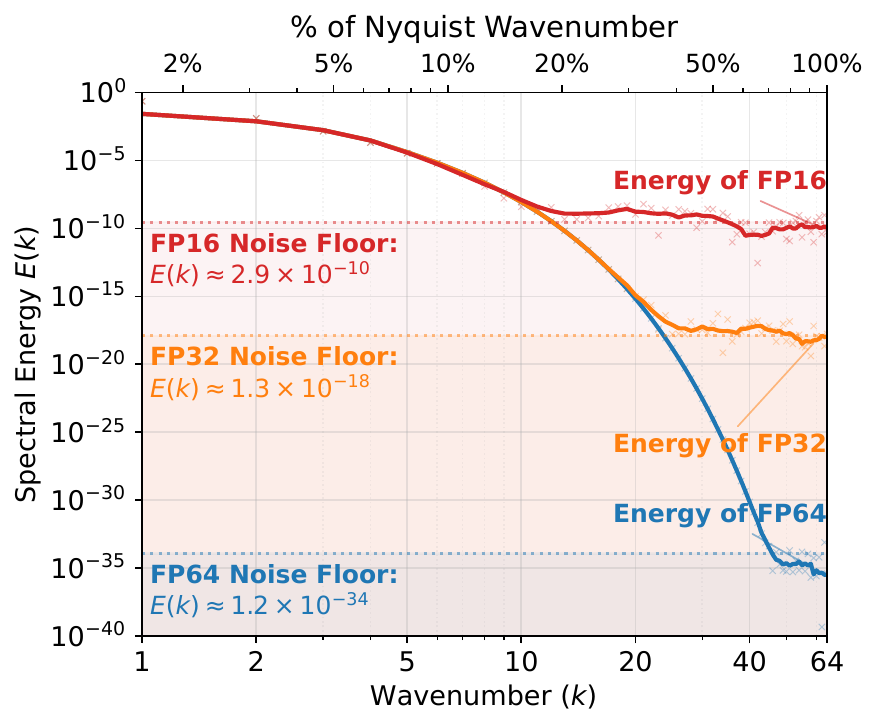}%
        \label{fig:spectral-128}%
    }\hfill
    \subfloat[Fine grid ($N_x=8192$)]{%
        \includegraphics[width=0.49\textwidth]{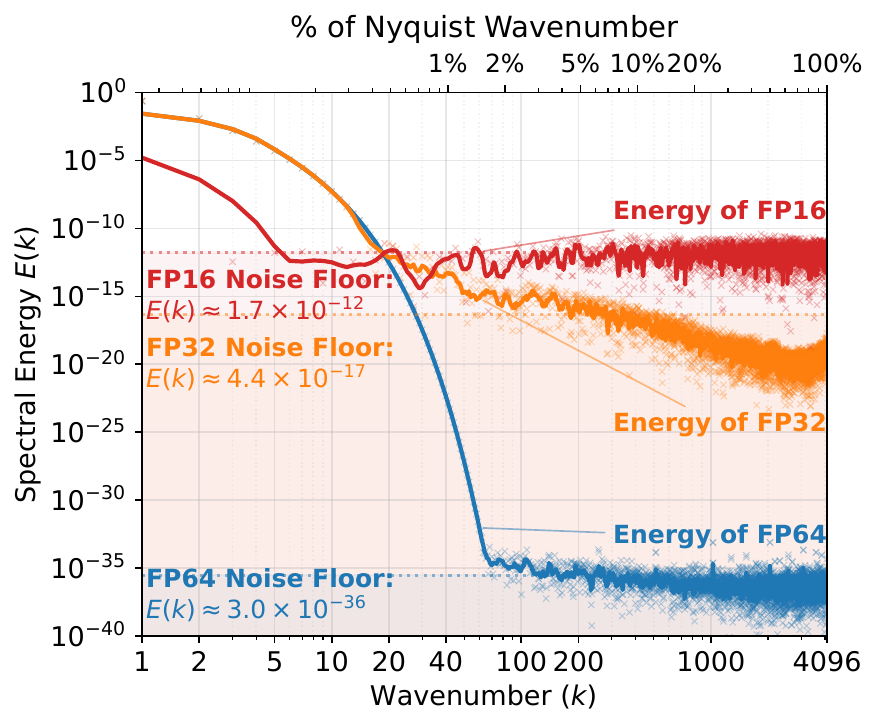}%
        \label{fig:spectral-8192}%
    }
    \caption{Spectral energy $E(k)$ of the velocity field (1D viscous Burgers', FTCS) at three precisions. Dotted lines and shading mark each precision's noise floor ($\propto \epsilon^2/N$). (a) For $N_x=128$, FP16 meets its noise floor near ${\sim}30\%$ of Nyquist and flattens, while FP32 and FP64 remain signal-dominated. (b) For $N_x=8192$, FP16 saturates near ${\sim}1\%$ of Nyquist, triggering a ``tail curl'' that feeds artificial energy back into macroscopic scales.}
    \label{fig:spectral-energy}
    \vspace{-10pt}
\end{figure*}

Standard floating-point tuning frameworks compare mixed-precision outputs against a high-precision reference (e.g., FP128) on the same grid~\cite{benz_dynamic_2012, demeure_algorithm_2023}. Under this methodology, switching from FP64 to FP32 reveals a sharp SNR drop---on the order of $180$\,dB---that appears catastrophic. But the drop is an artifact: because the reference shares the same grid, it inherits the same $E_\text{disc}$, so the SNR measures only the arithmetic gap between formats, not the physical accuracy of either~\cite{klower_low-precision_2021, palmer_more_2014}. Even with infinite precision the solver cannot exceed the discretization ceiling; extra significand bits merely reproduce truncation artifacts more faithfully.
 
\Cref{fig:perception-snr} confirms this. At practical resolutions ($h^{-1} \le 50$), the FP128-referenced curves (dashed) show a large FP64-to-FP32 gap, yet the analytical ones (solid) put both at the same ${\sim}68$\,dB floor. Only FP16 falls below it, a genuine precision limitation. \Cref{subsec:empirical-estimation-refinement} estimates this floor without $u^*$.

\subsection{Spectral Analysis and Stability}

Noise and signal are best separated in the frequency domain. By Parseval's theorem, the spatial variance $\text{Var}[u]$ equals the integral of the power spectral density $E(k)$ up to the Nyquist limit $k_{\text{max}} = \pi/h$~\cites[Ch.~12]{widrow_quantization_2008}[Ch.~9]{leveque_finite_2007}:
\begin{equation}
    \text{Var}[u] = \int_{0}^{k_{\text{max}}} E(k) \, dk.
\end{equation}
Round-off error, modeled as white noise, has a flat power spectral density $E_{\text{noise}}$. The accumulated variance $\text{Var}[E_{\text{ro}}] \propto \epsilon^2 h^{-2m}$ is distributed over the $d$-dimensional spectral volume $V_k \propto h^{-d}$, so the noise floor scales as $E_{\text{noise}}(\epsilon, h) \propto \epsilon^2 h^{d-2m}$.

Let $E_{\text{phys}}(k)$ be the power spectral density of the exact solution $u^*$. Physical signals conventionally exhibit power-law spectral decay in multiscale dynamical systems~\cite{kolmogorov_local_1991, nastrom_climatology_1985, skamarock_evaluating_2004}. For such signals characterized by a spectral decay rate $\alpha > 0$ such that $E_{\text{phys}}(k) \propto k^{-\alpha}$, the intersection with the noise floor defines a critical wavenumber $k_{\text{crit}} \propto (\epsilon^2 h^{d-2m})^{-1/\alpha}$. Beyond $k_{\text{crit}}$, the spectrum saturates at the spectral noise level, representing an unphysical energy accumulation at the grid scale. However, empirical evidence demonstrates that this elevated noise floor remains isolated at the truncation limit without interacting with or prematurely degrading the macroscopic scales~\cite{karp_effects_2025}.

\Cref{fig:spectral-energy} visualizes this threshold in the frequency domain. The physical energy cascade $E(k)$ decays at high wavenumbers, while floating-point round-off imposes a flat spectral noise floor $\propto \epsilon^2/N_x$~\cite[Ch.~12, 20]{widrow_quantization_2008}. Where that floor meets the cascade decides whether the noise is harmless. At coarse resolution it lands where viscous damping has already removed the energy. At fine resolution it intersects the cascade before dissipation sets in, and nonlinear interactions carry the grid-scale excess back into the resolved physical modes~\cite{zangl_icon_2015, palmer_towards_2012, karp_effects_2025}. This confirms that precision can be safely reduced whenever the solver operates in the discretization-limited regime.

\subsection{Empirical Estimation via Refinement}\label{subsec:empirical-estimation-refinement}

Because the exact analytical solution $u^*$ is unavailable for complex PDEs~\cite[Ch.~1]{leveque_finite_2007}, the true physical noise floor ($\text{SNR}_{\text{true}}$) cannot be computed directly. Instead, we estimate this limit empirically via full grid refinement, and show that temporal-only refinement provides a strict, practical over-estimate that bypasses some challenges of spatial refinement.

Given spatial and temporal orders of accuracy $p$ and $q$, the true discretization error $E_{\text{disc}}$ in the asymptotic regime expands as the sum of leading-order truncation terms:
\begin{equation}\label{eq:truncation_expansion}
    E_{\text{disc}}(h, \tau) = \tilde{u}_{h, \tau} - u^* = C_x h^p + C_t \tau^q + \mathcal{O}(h^{p+1}, \tau^{q+1}),
\end{equation}
where $C_x$ and $C_t$ are problem-dependent coefficients. By evaluating the solver at the target discretization $(h, \tau)$ and comparing the resulting state with a proportionally refined run $(h/r_x, \tau/r_t)$, we isolate these leading-order terms in a computable difference field~\cite[App.~A]{leveque_finite_2007}:
\begin{equation}\label{eq:richardson_difference}
    \begin{aligned}
    \Delta \tilde{u}_{h, \tau} &= \tilde{u}_{h, \tau} - \tilde{u}_{h/r_x, \tau/r_t} \\
    &\approx C_x h^p \left(1 - \frac{1}{r_x^p}\right) + C_t \tau^q \left(1 - \frac{1}{r_t^q}\right).
    \end{aligned}
\end{equation}

Because the refinement factors satisfy $r_x, r_t > 1$,  the variance of this computable difference field underestimates the true discretization error variance ($\text{Var}(\Delta \tilde{u}_{h, \tau}) < \text{Var}(E_{\text{disc}})$). This proxy thus yields a strictly conservative, upper-bound estimate of the grid's noise floor:
\begin{equation}\label{eq:snr_grid}
    \text{SNR}_{\text{grid}} = \frac{\text{Var}(\tilde{u}_{h, \tau})}{\text{Var}(\Delta \tilde{u}_{h, \tau})} > \frac{\text{Var}(\tilde{u}_{h, \tau})}{\text{Var}(E_{\text{disc}})} = \text{SNR}_{\text{true}}.
\end{equation}

\begin{figure}[!tb]
    \centering
    \includegraphics[width=\columnwidth]{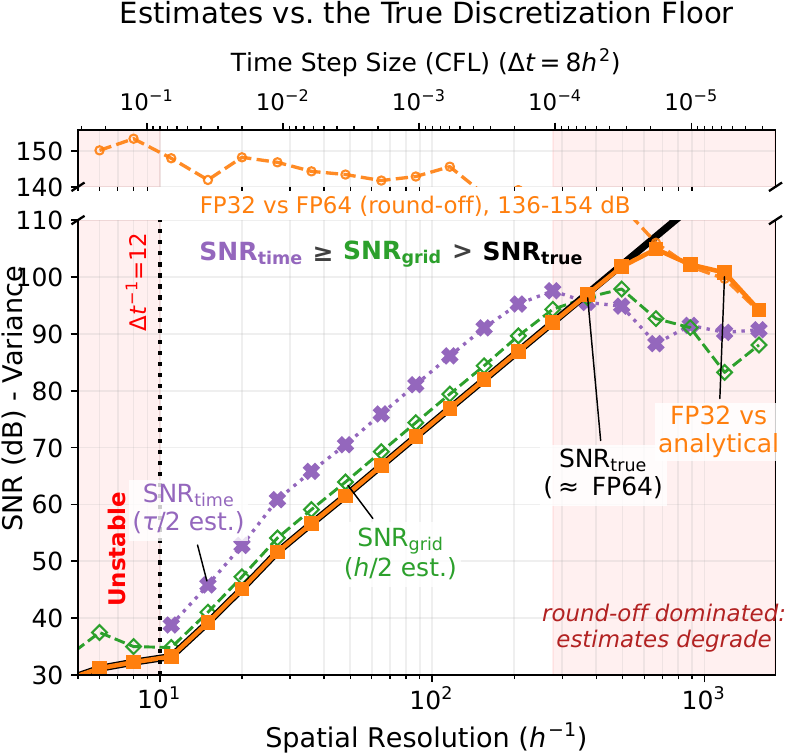}
    \caption{Floor estimates for the 1D viscous Burgers' equation. The temporal ($\tau/2$) and spatial ($h/2$) estimates bound the true floor from above across the discretization-limited range, confirming $\text{SNR}_{\text{time}} \ge \text{SNR}_{\text{grid}} > \text{SNR}_{\text{true}}$. Both degrade beyond $h^{-1} \approx 300$, where the difference fields are themselves round-off dominated. The top strip is the same-grid FP32-vs-FP64 comparison, off-scale at 136--154\,dB: an arithmetic reference registers none of this structure.}
    \label{fig:snr-chain}
    \vspace{-10pt}
\end{figure}

While $\text{SNR}_{\text{grid}}$ provides a sound threshold, full spatial refinement in large, complex models (e.g., 3D Earth systems) is often impossible when relying on empirical real-world data. Conversely, restricting refinement entirely to the temporal domain ($r_t > 1$, $r_x = 1$) is trivial in time-marching schemes:
\begin{equation}\label{eq:temporal_difference}
    \Delta \tilde{u}_{\tau} = \tilde{u}_{h, \tau} - \tilde{u}_{h, \tau/r_t} \approx C_t \tau^q \left(1 - \frac{1}{r_t^q}\right).
\end{equation}
Because stability constraints tie $\tau$ to $h$, $\Delta \tilde{u}_{\tau}$ captures only a fraction of the total discretization error ($\text{Var}(\Delta \tilde{u}_{\tau}) \le \text{Var}(\Delta \tilde{u}_{h, \tau})$), establishing a stricter noise floor:
\begin{equation}\label{eq:snr_time}
    \text{SNR}_{\text{time}} = \frac{\text{Var}(\tilde{u}_{h, \tau})}{\text{Var}(\Delta \tilde{u}_t)} \ge \text{SNR}_{\text{grid}} > \text{SNR}_{\text{true}}.
\end{equation}
\Cref{fig:snr-chain} validates this ordering on the 1D Burgers' equation, where the analytical $u^*$ makes $\text{SNR}_{\text{true}}$ computable. As \Cref{sec:performance-evaluation} shows, justifying the precision-lowering to FP32 suffices in practice. Since we only need to differentiate broad precision tiers (e.g., FP64 vs. FP32), this trivial temporal over-estimate is a highly practical, black-box criterion.

Therefore, our fundamental criterion for safe precision reduction dictates that noise from precision reduction must not degrade the signal below the full grid noise floor ($\text{SNR}_{\text{grid}}$). In practice, however, this operative threshold is satisfied by enforcing our stricter temporal over-estimate:
\begin{equation}\label{eq:safe_lowering_condition_time}
    \text{SNR}_{\text{var}}(\tilde{u}_{h, \tau}, \hat{u}_{h, \tau}) \ge \text{SNR}_{\text{time}}.
\end{equation}

\begin{figure*}[!tb]
    \centering
    \includegraphics[width=\linewidth]{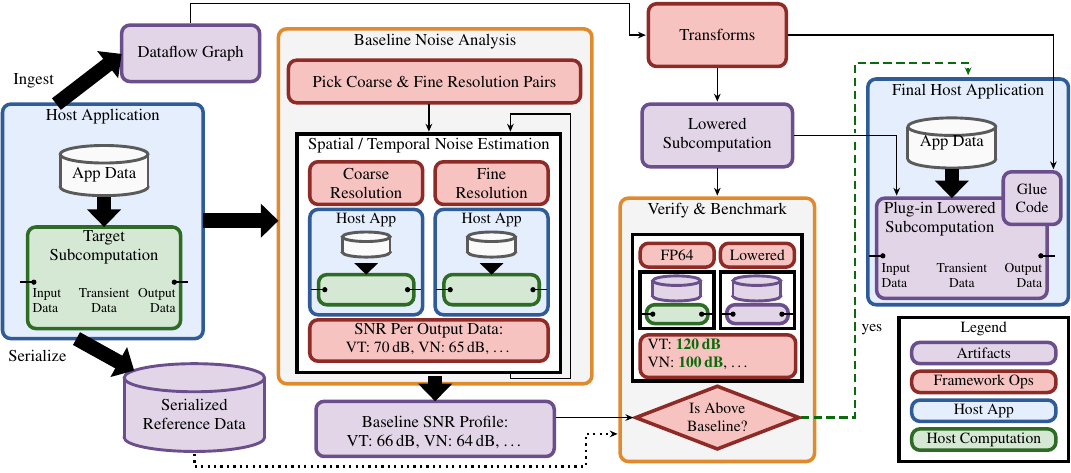}
    \caption{Automated Precision Lowering Workflow. A target subcomputation is isolated and translated into the DaCe IR alongside serialized reference data. A baseline noise analysis establishes the discretization noise floor using spatial or temporal refinement. The framework then uniformly downcasts the target subgraph to the desired precision and verifies that the resulting output SNR remains strictly above this baseline. The optimized kernel is subsequently wrapped in automatically generated glue code and reintegrated into the surrounding application.}
    \label{fig:workflow}
    \vspace{-10pt}
\end{figure*}

\section{Automated Precision Reduction Framework}\label{sec:automated-precision-lowering}

While \Cref{sec:discretization-noise-and-precision-limits} establishes the theoretical viability of precision lowering, manually downcasting complex scientific codebases is labor-intensive and error-prone~\cite{higham_mixed_2022}. Automating this requires a compilation framework that exposes an intermediate representation (IR) separating control and dataflow. We utilize the Data-Centric (DaCe) framework~\cite{ben-nun_stateful_2020}, whose dataflow IR provides the required granularity. DaCe represents programs as dataflow graphs, allowing us to target specific data containers for precision reduction---including intermediate transients that are invisible at the application interface. By modifying the types of these containers, dataflow analysis automatically tracks the downstream impact, enabling targeted precision reduction within specific computational subgraphs without invasive source code rewrites~\cite{klocke_computing_2025, bauer_digital_2021}.
 
This automated pipeline (\Cref{fig:workflow}) begins by extracting the target subcomputation from the surrounding application and translating it into the DaCe IR. It first runs a baseline noise analysis on coarse- and fine-resolution pairs, computing the discretization noise floor ($\text{SNR}_{\text{grid}}$, or $\text{SNR}_{\text{time}}$ where spatial refinement is unavailable) for each output container.
 
With this objective baseline established, the framework uniformly downcasts the entire dataflow subgraph to the target precision, wherever permitted by the compiler, and executes a verification pass against the serialized reference. We deliberately choose uniform over per-variable mixed precision: the coarse granularity avoids the combinatorial search over individual containers and scales to large codebases, at the cost of optimality (\Cref{sec:discussion-and-future-work}). If the noise of the lowered subgraph does not degrade the output SNR below the objective baseline, the target precision is validated as safe. Otherwise, the framework reports the failing fields and their SNR deficits, and the user can accept the baseline precision or investigate at finer granularity. The framework then compiles the optimized subgraph and generates the ``glue code'' for reintegration, so the host application needs no source-code changes.

Offloading to the GPU is another transformation on the same graph (\Cref{fig:workflow}), applied alongside the downcast but orthogonal to it. The framework itself runs once, offline: a few baseline and verification runs plus the lowering compilation. Its instrumentation adds a one-time initialization to the deployed run, amortized across timesteps. Beyond the boundary casts of \Cref{subsec:operation-semantics}, there is no per-timestep cost.

\subsection{Type Semantics and Application Interfacing}\label{subsec:operation-semantics}

Transforming a dataflow graph requires adapting arithmetic operations to modified storage types. Because the DaCe IR maps to C++, the framework leverages standard C++ type promotion rules rather than statically forcing operation types. The downcast targets all floating-point data containers---including intermediate accumulators and reduction temporaries---but leaves integer loop indices and control variables unchanged. Operations with fully narrowed inputs execute natively at the reduced precision ($\epsilon_{\text{low}}$). In mixed-precision scenarios involving both base ($\epsilon_{\text{base}}$) and lowered ($\epsilon_{\text{low}}$) operands, implicit widening to $\epsilon_{\text{base}}$ preserves intermediate accuracy and mitigates significance loss at internal precision boundaries.
 
If the entire application is captured within the IR, precision transitions are handled by the compiler. However, when lowering an isolated subgraph that must execute within a surrounding $\epsilon_{\text{base}}$ application, the framework must carefully manage the data interface. Typically, this requires explicit boundary casts: inputs are narrowed on ingestion ($\hat{x} = \text{cast}(x, \epsilon_{\text{low}})$), outputs widened on return ($y = \text{cast}(\hat{y}, \epsilon_{\text{base}})$). This isolates the subgraph's noise as a bounded perturbation, leaving the host's memory state intact.

Crucially, these boundary casts are not always necessary. Dataflow analysis detects when a lowered container is purely internal to the subgraph (e.g., a transient scratchpad buffer), eliminating the casting overhead. However, when boundary casts are unavoidable, they add per-invocation overhead at the subgraph interface. Each such cast is one streaming pass over a boundary array. These passes are bandwidth-bound, comparable to a copy kernel. Widening the lowered scope---ultimately to the full application---eliminates these transitions entirely. The overhead is therefore an artifact of lowering a region in isolation, not of the method itself. Note that analogous promotion overhead arises when the target hardware lacks native support for specific operations at $\epsilon_{\text{low}}$ (e.g., FP16 transcendentals); this effect is discussed in \Cref{sec:discussion-and-future-work}.

\section{Evaluation}\label{sec:performance-evaluation}

This section evaluates automated precision lowering by addressing two questions: (1) Can precision be reduced without compromising physical fidelity? and (2) Does the safe precision tier yield meaningful performance gains? We first establish the discretization noise floors of our target applications and evaluate numerical accuracy across the FP64, FP32, FP16 and BF16 formats. We then quantify the execution performance of the validated configurations, and close by comparing the criterion against the fixed tolerances and sensitivity bounds that existing tools require.

\subsection{Experiment Setup}\label{sec:experiment-setup}

\subsubsection{Applications}\label{subsec:setup-applications}

We evaluate two operational weather model components: Velocity Tendencies (VT) from the ICON dynamical core~\cite{zangl_icon_2015}, and the ECMWF CLOUDSC microphysics scheme~\cite{ecmwf_ecmwf-ifsdwarf-p-cloudsc_2025}. These represent the primary computational motifs in Earth-system modeling: memory-bound unstructured grid stencils (VT) and compute-bound column physics (CLOUDSC). VT is targeted specifically because it constitutes a dominant performance bottleneck within ICON.

VT evaluates the advective terms of the momentum equation in ICON's non-hydrostatic dynamical core, written in vector-invariant form: the kinetic-energy gradient, the vorticity flux, and vertical advection. It also computes the wind diagnostics these terms require. On the unstructured mesh, every such operator becomes an indirect stencil. ICON labels that mesh R$n$B$k$. Each edge of the base icosahedron is split into $n$ segments, and the resulting triangles are recursively bisected $k$ times, giving $20n^24^k$ cells. Every increment of $k$ therefore halves the mean cell width and quadruples the cell count. We evaluate R02B03--R02B06 (320, 160, 80 and 40\,km; \num{5120} to \num{327680} cells) at 90 vertical levels. VT is ingested as an SDFG and processed through the full automated pipeline (\Cref{sec:automated-precision-lowering}), serving as the end-to-end demonstration of framework-driven precision lowering.

CLOUDSC parameterizes cloud microphysics. It advances temperature, specific humidity, cloud fraction and the condensate types---liquid, ice, rain and snow---through the conversions among them: condensation, autoconversion, accretion, freezing, melting and sedimentation. Vertical columns are independent. We evaluate it in its existing single-precision build mode, which validates the SNR methodology on a second, structurally distinct application. The kernel advances one physics timestep ($t_{\mathrm{phys}} = 120$\,s) by forward Euler integration, split into $N$ equal steps. Performance runs use $N=10$; the SNR evolution of \Cref{fig:snr-cloudsc} uses $N=\num{10000}$, integrating the same 120\,s at a $1000\times$ smaller step.

\subsubsection{System and Compilation}\label{subsec:setup-hardware}

Experiments are conducted on single nodes of two CSCS systems: a node of the Alps system equipped with an NVIDIA GH200 Grace Hopper Superchip (96\,GB HBM3, 4\,TB/s peak bandwidth)~\cite{fusco_understanding_2024} and CUDA 12.6; and an Ault node equipped with an NVIDIA A100 SXM4 (40\,GB HBM2e, 1.56\,TB/s peak bandwidth) and CUDA 12.9.

Applications are compiled into shared libraries from DaCe-generated CUDA code using \texttt{nvcc -O3 --use\_fast\_math}, targeting \texttt{sm\_90} on GH200 and \texttt{sm\_80} on A100. Precision lowering, GPU offloading, and boundary cast insertions are managed by the automated DaCe pipeline (\Cref{sec:automated-precision-lowering}). Reproducibility scripts and configuration details are in the Artifact Description appendix.

\subsubsection{Execution and Methodology}\label{subsec:setup-methodology}

All benchmarking follows established best practices for reproducible performance reporting~\cite{hoefler_scientific_2015}. Hardware profiling (e.g., bandwidth, utilization) is collected on GH200 via \texttt{ncu --set full} in a standalone test harness that executes a single computational step from a serialized input snapshot, isolating the kernel from host framework interference; warmup iterations are excluded from profiling. CLOUDSC wall-clock timings are measured directly within this harness, while ICON-VT timings are extracted from full model runs due to tight coupling with the dynamical core. All timings exclude boundary casting overhead (i.e., FP64-to-FP32/FP16/BF16 interface conversions).

Numerical quality is quantified by the variance SNR relative to the FP64 baseline. A target precision is deemed safe only if its SNR remains above the discretization noise floor (\Cref{subsec:suitability-precision-reduction}). For ICON-VT, this baseline is estimated via temporal refinement ($\text{SNR}_\text{time}$); for CLOUDSC, temporal ($\text{SNR}_\text{time}$) and vertical ($\text{SNR}_\text{grid}$) refinement baselines are computed.

\subsection{Numerical Accuracy}\label{subsec:suitability-precision-reduction}

\begin{figure}[!tb]
\centering
\subfloat[ICON-VT: vn SNR over 25k physics steps (160\,km, $\tau_{\mathrm{phys}}=8$\,s). FP32 (${\sim}50-154$\,dB) stays above the temporal floor (${\sim}25-62$\,dB) throughout. Its sharp drop near step ${\sim}200$ marks the point where roundoff saturates, well above the floor. FP16 (${\sim}3-76$\,dB) falls below the floor within ${\sim}200$ steps, and BF16 (${\sim}3-58$\,dB) starts at it.]{%
  \includegraphics[width=\linewidth]{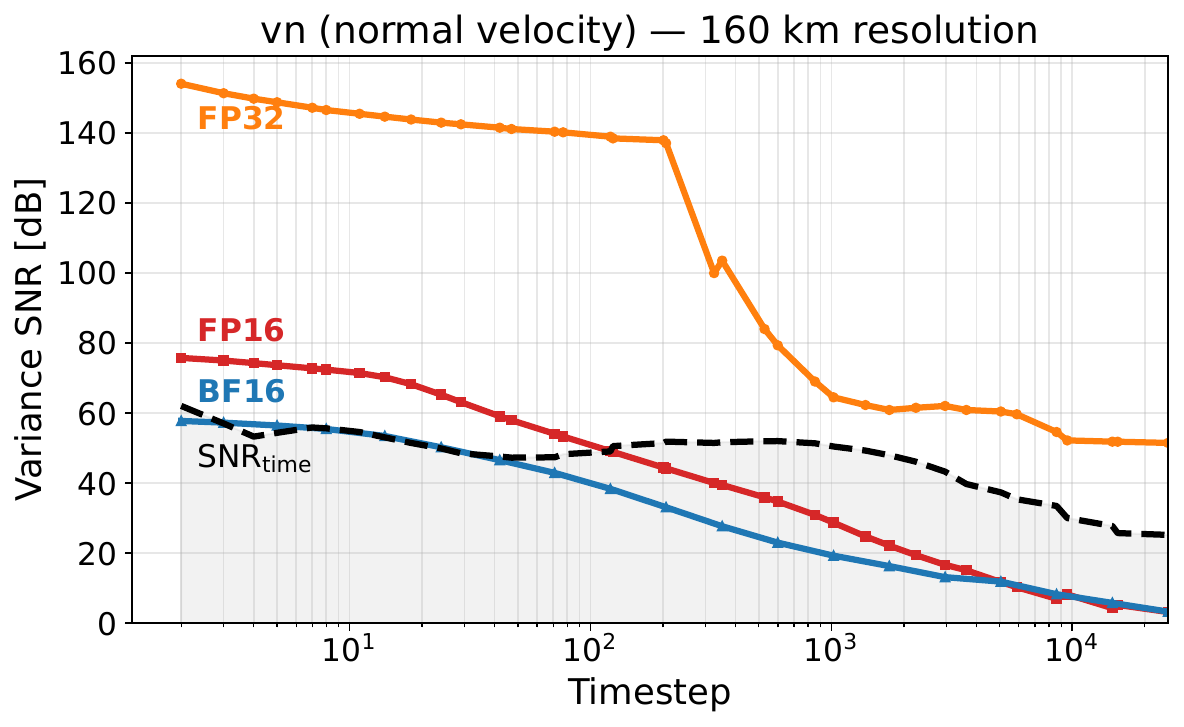}%
  \label{fig:snr-vt}%
}\hfill
\subfloat[CLOUDSC: PT SNR over \num{10000} steps for a 120\,s interval ($\Delta t = 12$\,ms). FP32 (${\sim}65-128$\,dB) starts just above the temporal floor $\mathrm{SNR_{time}}$ (${\sim}35-125$\,dB) and pulls ${\sim}30$\,dB clear as that floor falls. The spatial floor $\mathrm{SNR_{grid}}$ (${\sim}32-77$\,dB) lies below both: flat while profiles are smooth, then dropping sharply when a cloud layer forms (${\sim}$step $100-300$).]{%
  \includegraphics[width=\linewidth]{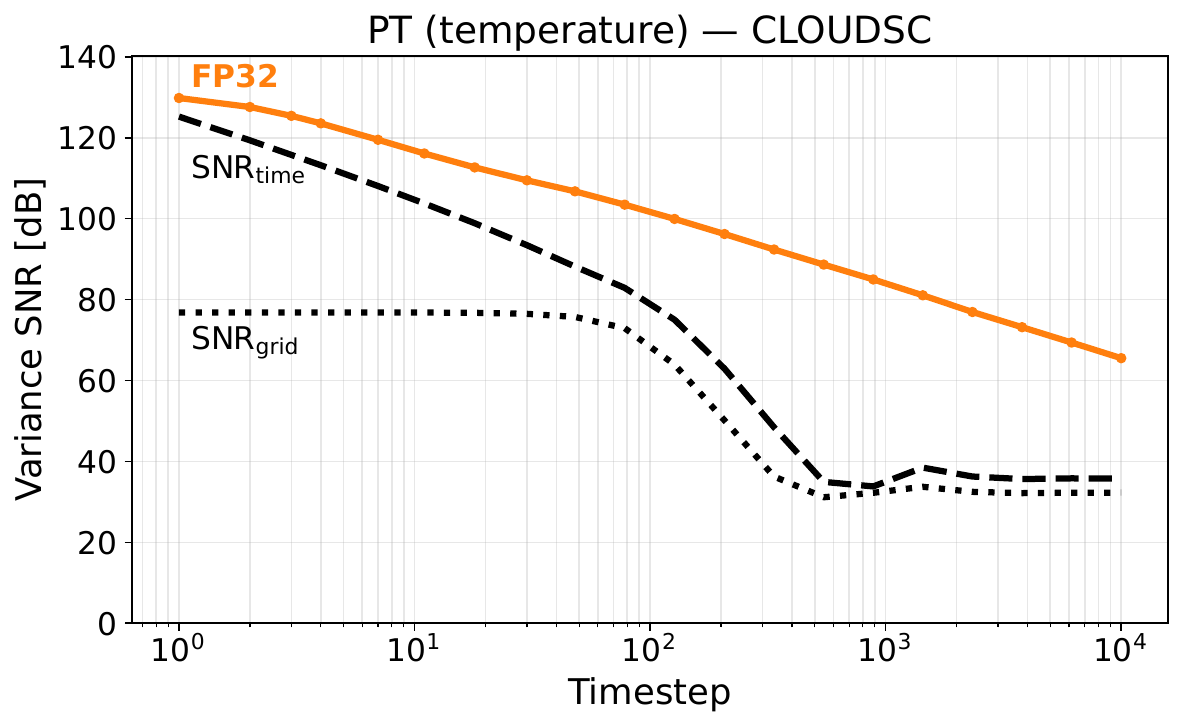}%
  \label{fig:snr-cloudsc}%
}
\caption{$\text{SNR}_\text{var}$ evolution over the full runs. In both applications the floors fall as the step count grows, and FP32 stays clear throughout.}
\label{fig:temporal-noise}
\vspace{-10pt}
\end{figure}

\begin{table}[!tb]
\centering
\caption{\normalfont\small Per-field variance SNR [dB] for ICON velocity tendencies precision modes vs.\ FP64 baseline at four grid resolutions, after $p{=}2$ physics timesteps. Underlined entries indicate that the reduced-precision mode falls below $\text{SNR}_{\text{time}}$ at that resolution (i.e., no longer viable).}\label{tab:icon-vt-snr-results}
\begingroup
\small
\setlength{\tabcolsep}{4pt}
\renewcommand{\arraystretch}{1.05}
\begin{tabularx}{0.95\columnwidth}{@{}l
  >{\raggedleft\arraybackslash\hsize=0.9\hsize}X
  >{\raggedleft\arraybackslash\hsize=0.9\hsize}X
  >{\raggedleft\arraybackslash\hsize=0.9\hsize}X
  >{\raggedleft\arraybackslash\hsize=0.9\hsize}X
  >{\raggedleft\arraybackslash\hsize=1.2\hsize}X@{}}
\toprule
& vn & w & vt & vn\_ie & w\_concorr\_c \\
\midrule
\multicolumn{6}{@{}c@{}}{320 km Resolution} \\
\midrule
FP32 & 154.0 & 139.7 & 151.2 & 151.0 & 151.4 \\
FP16 &  75.7 &  61.8 &  73.0 &  72.8 &  82.3 \\
BF16 &  \underline{57.8} &  44.4 &  \underline{54.9} &  \underline{54.9} &  64.8 \\
$\text{SNR}_\text{time}$ & 58.8 & 31.1 & 65.3 & 61.5 & 58.7 \\
\addlinespace[2pt]
\midrule
\multicolumn{6}{@{}c@{}}{160 km Resolution} \\
\midrule
FP32 & 154.1 & 135.4 & 151.3 & 151.1 & 151.4 \\
FP16 &  75.8 &  57.4 &  73.0 &  72.9 &  82.7 \\
BF16 &  \underline{57.8} &  39.9 &  \underline{55.0} &  \underline{55.0} &  \underline{64.8} \\
$\text{SNR}_\text{time}$ & 66.6 & 30.4 & 72.4 & 69.2 & 69.9 \\
\addlinespace[2pt]
\midrule
\multicolumn{6}{@{}c@{}}{80 km Resolution} \\
\midrule
FP32 & 154.1 & 131.6 & 151.3 & 151.1 & 151.4 \\
FP16 &  75.8 &  53.6 &  \underline{73.0} &  \underline{73.0} &  82.8 \\
BF16 &  \underline{57.8} &  36.2 &  \underline{55.0} &  \underline{55.1} &  \underline{64.5} \\
$\text{SNR}_\text{time}$ & 73.2 & 30.2 & 79.0 & 76.5 &  77.0 \\
\addlinespace[2pt]
\midrule
\multicolumn{6}{@{}c@{}}{40 km Resolution} \\
\midrule
FP32 & 154.1 & 126.9 & 151.3 & 151.1 & 151.4 \\
FP16 &  \underline{75.8} &  48.7 &  \underline{73.0} &  \underline{73.0} &  \underline{82.7} \\
BF16 &  \underline{57.8} &  31.4 &  \underline{55.0} &  \underline{55.1} &  \underline{64.6} \\
$\text{SNR}_\text{time}$ & 80.9 & 30.1 & 86.9 & 83.8 & 85.1 \\
\bottomrule
\end{tabularx}
\endgroup
\end{table}

\begin{table}[!tb]
\centering
\caption{\normalfont\small Per-field variance SNR [dB] for CLOUDSC precision modes vs.\ FP64 baseline at $p{=}2$, $t_{\mathrm{phys}} = 120$\,s. Values are independent of column count (164k--656k) as CLOUDSC operates column-wise on identical input. Temporal discretization reference from nsub\,=\,1 vs.\ 2; grid reference from $k_{\mathrm{lev}}$ 137 vs.\ 274 with nsub\,=\,1 vs.\ 2, refining both discretizations together. ``--'' indicates divergence to NaN.}\label{tab:cloudsc-snr-results}
\begingroup
\small
\setlength{\tabcolsep}{4pt}
\renewcommand{\arraystretch}{1.05}
\begin{tabular}{@{}lrrrr@{}}
\toprule
Field & FP32 & FP16 & $\mathrm{SNR}_\mathrm{time}$ & $\mathrm{SNR}_\mathrm{grid}$ \\
\midrule
PT   & 124.2 &   -- & 112.3 & 46.4 \\
PQ   & 116.2 &   -- &  92.1 & 45.1 \\
PA   &  93.5 &  3.8 &  75.8 & 26.3 \\
PCLV &  81.4 & -0.2 &  12.0 & 10.4 \\
\bottomrule
\end{tabular}
\endgroup
\end{table}

\begin{table}[!tb]
\centering
\begin{threeparttable}
\caption{\normalfont\small \texttt{ncu} profiling for ICON velocity tendencies (VT) on GH200, warm rep; cold and warm differ by $<\!5\%$ on all metrics. Wall-clock times are in \Cref{fig:kernel-bars} --- \texttt{ncu} locks clocks to base frequencies, so its timings are not production-comparable.}\label{tab:ncu-profiling}
\begingroup
\small
\renewcommand{\arraystretch}{1.10}

\begin{tabular*}{\columnwidth}{@{\extracolsep{\fill}}lrrrrrr@{}}
\toprule
 & \multicolumn{3}{c}{Memory} & \multicolumn{3}{c}{Compute share (\%)\tnote{\ddag}} \\
\cmidrule(lr){2-4} \cmidrule(lr){5-7}
FP & \makecell{Read\\(MB)} & \makecell{Write\\(MB)\tnote{\dag}} & \makecell{BW util.\\(\%)} & FP64 & FP32 & \makecell{FP16/\\BF16} \\
\midrule
\multicolumn{7}{@{}c@{}}{320 km Resolution} \\
\midrule
FP64 & 118  & 0.1  & 27.2 & 95.2 & 2.9  & 1.9 \\
FP32 & 60.0 & 0.0  & 15.7 & 0    & 91.0 & 9.0 \\
FP16 & 31.4 & 0.0  & 8.2  & 0    & 13.5 & 86.5 \\
BF16 & 31.4 & 0.0  & 8.2  & 0    & 27.3 & 72.7 \\
\addlinespace[2pt]
\midrule
\multicolumn{7}{@{}c@{}}{160 km Resolution} \\
\midrule
FP64 & 461 & 45.8 & 51.8 & 95.2 & 2.9  & 1.9 \\
FP32 & 233 & 8.1  & 31.3 & 0    & 92.0 & 8.0 \\
FP16 & 121 & 0.6  & 15.7 & 0    & 15.6 & 84.4 \\
BF16 & 121 & 0.3  & 15.4 & 0    & 28.0 & 72.0 \\
\addlinespace[2pt]
\midrule
\multicolumn{7}{@{}c@{}}{80 km Resolution} \\
\midrule
FP64 & 1{,}850 & 461  & 70.5 & 95.2 & 2.9  & 1.9 \\
FP32 & 931     & 187  & 47.3 & 0    & 92.0 & 8.0 \\
FP16 & 483     & 58.5 & 23.7 & 0    & 15.3 & 84.7 \\
BF16 & 483     & 58.9 & 22.9 & 0    & 27.9 & 72.1 \\
\addlinespace[2pt]
\midrule
\multicolumn{7}{@{}c@{}}{40 km Resolution} \\
\midrule
FP64 & 11{,}350 & 3{,}850 & 75.2 & 95.7 & 2.0  & 2.4 \\
FP32 & 5{,}220  & 1{,}860 & 51.6 & 0    & 93.5 & 6.5 \\
FP16 & 2{,}660  & 913     & 28.6 & 0    & 17.0 & 83.0 \\
BF16 & 2{,}660  & 915     & 27.7 & 0    & 28.5 & 71.5 \\
\bottomrule
\end{tabular*}

\begin{tablenotes}[flushleft]
  \footnotesize
  \item[\dag] DRAM write measures L2$~\to~$DRAM writebacks. At 320\,km the working set fits in L2 (96\,MB on GH200), so stores are absorbed entirely by L2 with no DRAM evictions during the kernel.
  \item[\ddag] FMA (fused multiply-add) counted as 2 FLops~\cite{nvidia_matrix_2023}.
\end{tablenotes}

\endgroup
\end{threeparttable}
\end{table}

We estimate each baseline noise level by comparing against refined configurations (\Cref{subsec:empirical-estimation-refinement}). While temporal refinement is straightforward for explicit schemes, spatial refinement is challenging for unstructured grids like ICON's. Because $\text{SNR}_\text{time}$ proves sufficient to confirm the viability of FP32 (which clearly preserves accuracy on all grids), and of FP16 on coarser grids, we rely on the temporal baseline for ICON-VT. Conversely, CLOUDSC's 1D column physics features a simple, horizontally invariant structure, permitting a straightforward vertical refinement to obtain a meaningful $\text{SNR}_\text{grid}$.
 
\Cref{fig:temporal-noise} tracks the variance SNR over successive timesteps for one prognostic variable from each application---\texttt{vn} for ICON-VT and \texttt{PT} for CLOUDSC---while \Cref{tab:icon-vt-snr-results} and \Cref{tab:cloudsc-snr-results} give it across all primary output fields. This profiling exposes prognostic fields that are inherently noise-dominated at the grid scale. For example, the vertical velocity (\texttt{w}) in ICON-VT and cloud condensate (\texttt{PCLV}) in CLOUDSC stabilize at floors of ${\sim}30$\,dB and ${\sim}12$\,dB even in FP64. Discretization noise already eclipses the lower-order significand bits for such variables, so a large error there is expected: a lowering is safe only if its output SNR stays above these baselines.
 
Evaluated against this criterion, FP32 is safe for both applications. In CLOUDSC, \texttt{PT} and \texttt{PQ} retain FP32 SNRs of 124.2\,dB and 116.2\,dB, comfortably above both their temporal (112.3\,dB, 92.1\,dB) and vertical grid baselines (46.4\,dB, 45.1\,dB). All evaluated ICON-VT fields clear their respective $\text{SNR}_\text{time}$ thresholds by wide margins in single precision.

FP16 presents a mixed picture. In CLOUDSC, uniform FP16 catastrophically diverges to NaN for \texttt{PT} and \texttt{PQ}, and its SNR for \texttt{PA} drops to 3.8\,dB---falling well below both discretization floors. In ICON-VT, a uniform FP16 pass identified \texttt{w\_concorr\_c} as the sole variable that diverges immediately when lowered; retaining it at FP32 while lowering all remaining containers to FP16 yields the configuration shown in \Cref{tab:icon-vt-snr-results}. All fields clear their $\text{SNR}_\text{time}$ thresholds at coarse resolutions, but the rising discretization floor overtakes FP16's noise level at 80\,km and beyond (underlined entries). These are two-step verdicts. \Cref{fig:snr-vt} shows that the coarse-grid margin is not durable: at 160 km, FP16 crosses the floor within ${\sim}200$ steps, so its admissibility here holds for short integrations only.

BF16 fares worse still. Across every field and resolution it sits a uniform ${\sim}18$\,dB below FP16---three fewer significand bits at ${\sim}6$\,dB. Only \texttt{w} remains above $\text{SNR}_\text{time}$ at all four resolutions, and \texttt{w\_concorr\_c} only at 320\,km.

\begin{figure*}[!tb]
\centering
\subfloat[ICON velocity tendencies (VT; aquaplanet) at four resolutions. Each dynamics substep evaluates a predictor and a corrector path; we report the predictor, as the corrector's speedups match. BF16 tracks FP16 to within $0.1\times$, as the two move identical memory traffic.]{%
  \includegraphics[width=0.584\textwidth]{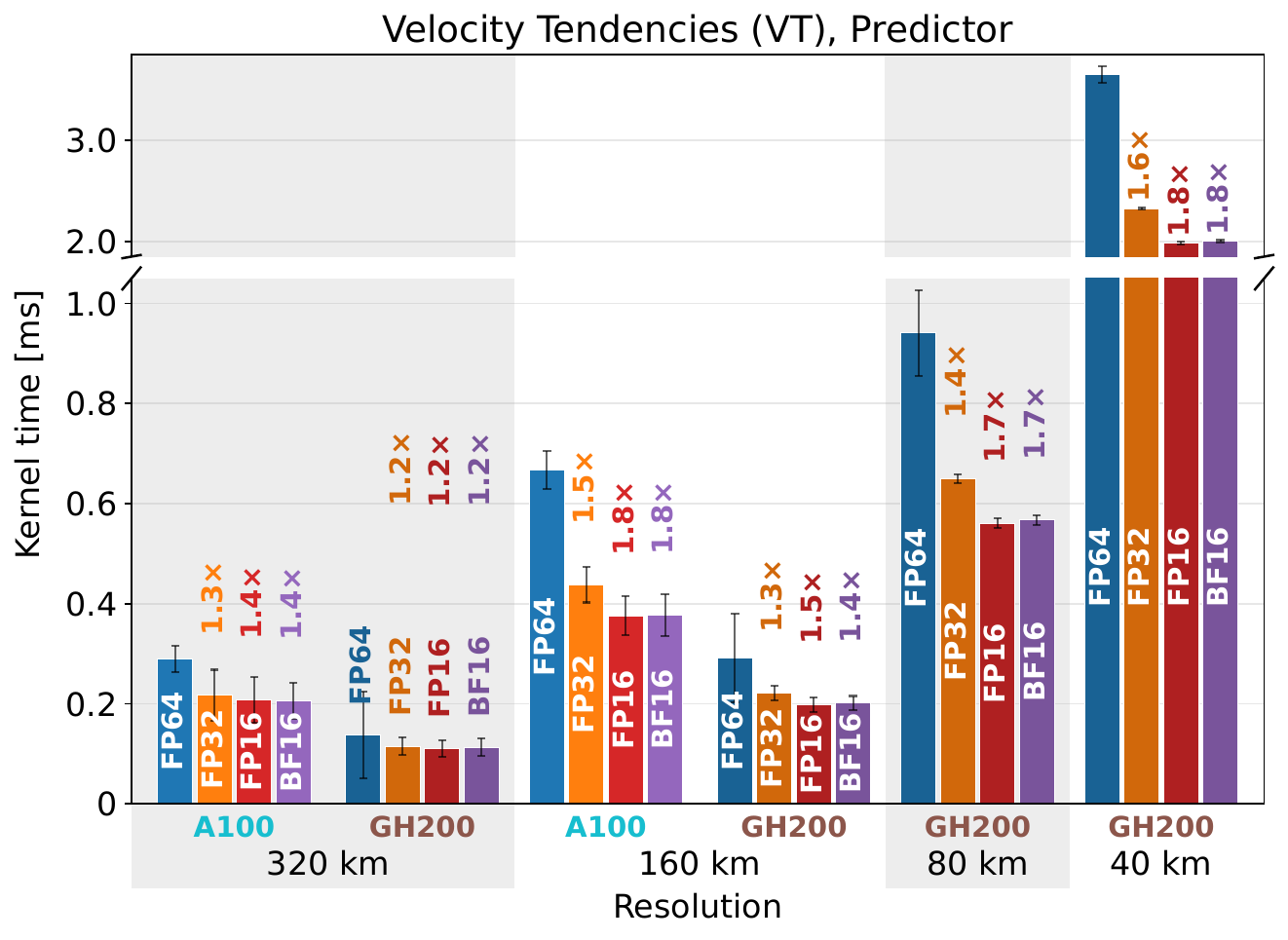}%
  \label{fig:kernel-bars:vt}%
}\hfill
\subfloat[CLOUDSC microphysics (41k--656k columns). FP16 is omitted due to numerical divergence (\Cref{subsec:suitability-precision-reduction}).]{%
  \includegraphics[width=0.396\textwidth]{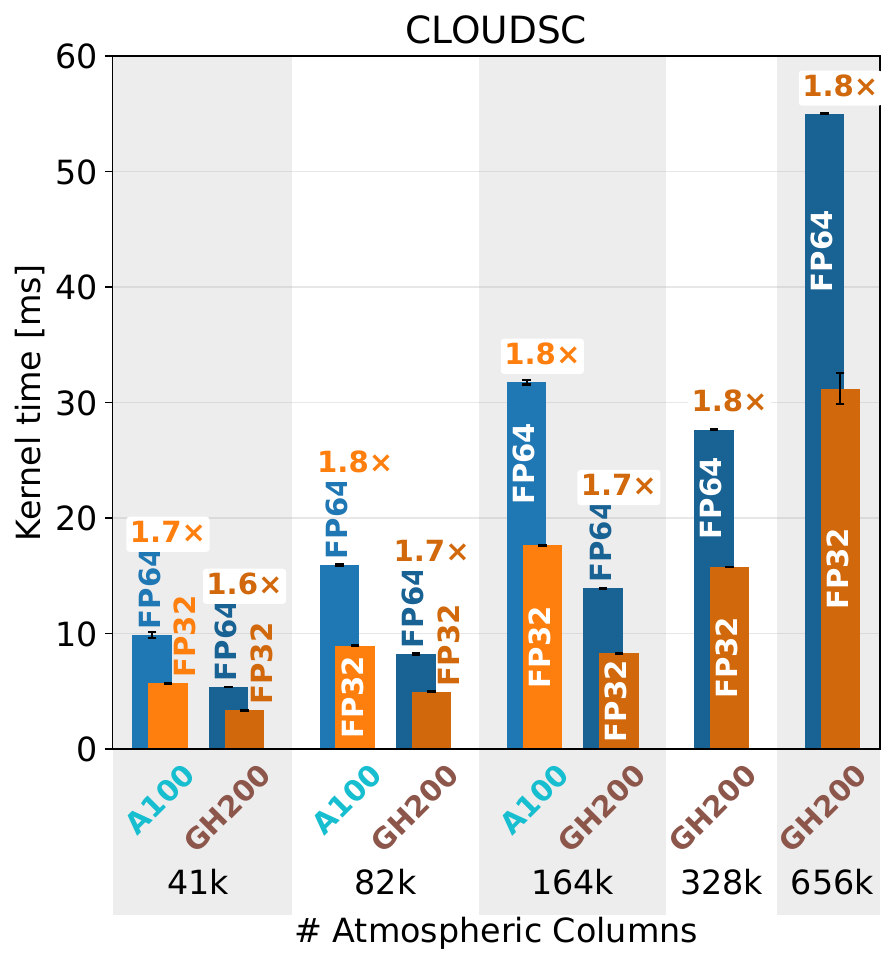}%
  \label{fig:kernel-bars:cloudsc}%
}
\caption{Host-side GPU kernel wall-clock time on A100 and GH200; all possible data lowered to the respective precision; speedup relative to FP64. A100 covers only the grids that fit its 40\,GB HBM (VT up to 160\,km; CLOUDSC up to 164k columns); larger grids are GH200-only.}%
\label{fig:kernel-bars}
\vspace{-8pt}
\end{figure*}

\subsection{Computational Performance}\label{subsec:computational-performance}

Having established that FP32 preserves numerical fidelity on both applications, while FP16 is safe only on the coarser VT grids and fails outright in CLOUDSC, we now quantify the resulting performance impact. \Cref{fig:kernel-bars} details execution times and relative speedups for the precision-lowered VT and CLOUDSC kernels. VT evaluates a predictor and a corrector path in each dynamics substep; we report the predictor, as the corrector shows the same speedup profile. Transitioning from FP64 to FP32 yields substantial performance improvements: CLOUDSC gains $1.6$--$1.8\times$ on GH200 and $1.7$--$1.8\times$ on A100, while VT rises from $1.2\times$ at 320\,km to $1.6\times$ at 40\,km on GH200, and reaches $1.3$--$1.5\times$ on A100. Because these kernels are memory-bandwidth-bound, halving the data element size halves DRAM traffic, allowing performance gains to approach the theoretical $2\times$ limit as the problem size grows.

For FP16, CLOUDSC is ruled out entirely. For VT, FP16 is the fastest configuration measured, reaching $1.8\times$ at 40\,km on GH200 and $1.8\times$ at 160\,km on A100, and BF16 tracks it to within $0.1\times$ everywhere, since the two move identical memory traffic. The advantage is resolution-dependent: at 320\,km both merely match FP32. The largest gains also fall outside the admissible window, as \Cref{tab:icon-vt-snr-results} clears FP16 only at 320 and 160\,km and BF16 only for \texttt{w}. At 160\,km the two coincide, with FP16 both admissible over short integrations and faster than FP32 ($1.5\times$ against $1.3\times$ on GH200).

Hardware profiling data (\Cref{tab:ncu-profiling}) confirms this plateau. For ICON-VT, downcasting from FP64 to FP32 halves DRAM reads (e.g., $1{,}850$\,MB to 931\,MB at 80\,km) and increases L1/L2 cache hit rates due to the compressed memory footprint. However, the transition from FP32 to FP16 is not matched by a comparable speedup despite DRAM traffic halving further to 483\,MB. Profiling reveals that as data volume shrinks, DRAM bandwidth utilization drops from 70.5\% at FP64 to 23.7\% at FP16 for 80\,km, and to 8.2\% for 320\,km. This indicates that at FP16, the kernel is no longer constrained by main memory bandwidth; consequently, further compressing the memory footprint provides no execution time benefit. The compute mix compounds this: whereas the FP64 build issues 95\% of its floating-point work in FP64 and the FP32 build 92\% in FP32, the FP16 build retains roughly a seventh at FP32, because the \texttt{w\_concorr\_c} chain (\Cref{subsec:suitability-precision-reduction}) underflows in half precision and is held there. BF16 moves FP16's memory traffic but retains nearly twice that share at FP32, since division and square root lack native bfloat16 instructions.

\subsection{Field-Dependent Thresholds}\label{subsec:criteria-comparison}

\begin{figure}[!t]
\centering
\subfloat[SNR of each output field against the FP64 run: one row per field, one marker per format emulated by RAPTOR. The green step is that field's floor, obtained by evaluating the kernel at $137$ and at $274$ vertical levels with the substep count held fixed. Triangles at the left edge mark fields whose lowered output was not finite; at the right edge, SNRs far above the floor, clipped there.]{%
  \includegraphics[width=\linewidth]{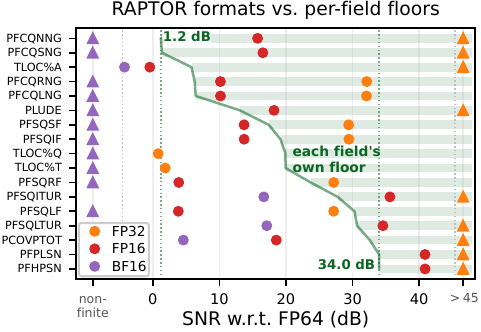}%
  \label{fig:cloudsc-floor}%
}\\
\subfloat[ADAPT is run once per output field, with that field as its sole dependent, and the error bound it is given is swept. The axis is that bound $|\Delta J|$ as a fraction of the field's squared norm $J$, in decibels, not the field-variance $\mathrm{SNR_{var}}$ of \Cref{eq:snr_var}. Each curve gives how many of the $55$ registered variables ADAPT then refuses to demote from FP64. Six fields are shown; the cross on a curve marks the bound refinement derives for that field, refining the vertical levels and the substep together as \Cref{eq:richardson_difference} requires.]{%
  \includegraphics[width=\linewidth]{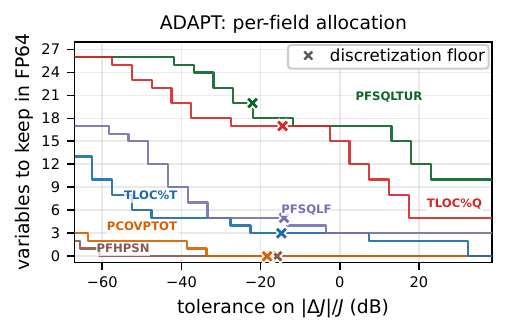}%
  \label{fig:cloudsc-adapt}%
}
\caption{A $120$\,s CLOUDSC physics interval, taken in two substeps. Each of the $17$ output fields for which a floor is defined sets its own threshold.}\label{fig:criteria}
\vspace{-10pt}
\end{figure}

We now contrast our method with two tools representing the state of the art in precision lowering: RAPTOR~\cite{hoerold_raptor_2025} and ADAPT~\cite{menon_adapt_2019}. The existing tools measure the effect of precision lowering, but leave the fundamental question of what deviation is acceptable open to the user: RAPTOR renders a lowered execution to inspect, ADAPT demands an error bound as input. We demonstrate that the underlying principle of our work addresses this gap of missing tolerance.

\paragraph*{RAPTOR} RAPTOR emulates a format by truncating every arithmetic operation. An LLVM instrumentation pass replaces floating-point operations in user-selected scopes (function, file, or program) with calls into a runtime that executes them via MPFR at a user-specified exponent and significand width, casting the result back to the original type after every operation. The tool leaves to the user the decision of what deviation from the full-precision run is tolerable. Furthermore, RAPTOR's truncation is scope-based (function, file, or program), while our dataflow-based approach permits finer-grained control over mixed precision (e.g., per data container).

We emulate FP32, FP16 and BF16 with RAPTOR and compare every output against the FP64 run in $\mathrm{SNR_{var}}$ (\Cref{fig:cloudsc-floor}). The SNRs improve with more significand bits wherever the lowered run stays finite, but the distribution differs from field to field. RAPTOR, at its best, can use a predefined threshold to declare safety of the lowering (i.e. equivalent to draw a vertical line on \Cref{fig:cloudsc-floor}). But how much error do we already accept at FP64? That becomes clearer only when we account for the estimated discretization floor, spanning from $1.2$ to $34.0$\,dB across $17$ lowered fields. FP32 falls below it for $3$ of them, FP16 for $8$, and BF16 for all $17$. We can conclude that those fields must remain in higher precision when lowering.

\paragraph*{ADAPT} ADAPT ranks variables by adjoint sensitivity against a given bound. The user specifies a scalar quantity of interest $J$ (e.g., norm of a field or residual), and a set of candidate variables for demotion from the working precision to a single lower one---FP64 to FP32 in practice. A single run under reverse-mode algorithmic differentiation yields, for every candidate $x$, the sensitivity $\partial J/\partial x$ at each point where $x$ is assigned. ADAPT combines these sensitivities with the rounding error the lower precision would introduce into $x$ to obtain a first-order estimate of the resulting error in $J$, and marks variables for demotion in order of increasing estimated impact until a user-specified bound on $\lvert\Delta J\rvert$ is exhausted. The bound is an absolute deviation of $J$ from the FP64 run. This is not compatible with the SNR approach we take, but the underlying principle is demonstrably useful even in this case.

We run ADAPT once per output field, with that field as its sole dependent, and sweep the error bound it is given (\Cref{fig:cloudsc-adapt}). Tightening the bound makes ADAPT keep more of the $55$ demotable arrays in FP64, and the curve is different for every field. As before, choosing where on that curve to sit is left to the user, and this is where our simple refinement gives an estimated bound on how much error is tolerable for every output. Those bounds span $24$\,dB across the $16$ fields swept, and at them ADAPT leaves no arrays in FP64 for \texttt{PCOVPTOT} and \texttt{PFHPSN} but $20$ for \texttt{PFSQLTUR}.

\section{Related Work}\label{sec:related-work}

\paragraph*{Reduced Precision in Weather and Climate}
Key atmospheric variables retain only 7--12 bits of real information~\cite{klower_low-precision_2021}, the ECMWF IFS runs in single precision with negligible forecast degradation~\cite{vana_single_2017}, and the freed budget is better spent on ensemble size or resolution~\cite{palmer_more_2014, karp_effects_2025}. These studies validate precision reduction \emph{a~posteriori}---through ensemble statistics or distributional comparison---but do not provide an automated, scalar criterion for a compiler-level tuning loop. Our work complements these findings with a principled threshold and a framework that automates the lowering.
 
\paragraph*{Automated Precision Tuning}
Dynamic approaches like PRECIMONIOUS~\cite{rubio-gonzalez_precimonious_2013} search for mixed-precision configurations satisfying user-defined error thresholds, but lack correctness guarantees for untested inputs. Static approaches like Daisy~\cite{darulova_daisy_2018} compute worst-case rounding bounds via abstract interpretation or SMT solvers, but these bounds are often impractically pessimistic for long-running PDE simulations where errors cancel probabilistically~\cite{isychev_cost_2025}. Compiler-integrated rewriting goes further: Poseidon~\cite{qian_thinking_2026} instruments the program in a first pass, and runs it to capture numerical context. Then it uses that profile in a second pass to synthesize candidate rewrites, solving for a Pareto set of performance/accuracy tradeoffs. Emulators such as RAPTOR~\cite{hoerold_raptor_2025} take a format rather than a tolerance, leaving the verdict to the user. These tools measure error against a same-grid high-precision reference. Our framework evaluates against the physical discretization floor instead~(\Cref{subsec:moving-beyond-arithmetic-baselines}). Discretization-error-accurate multigrid~\cite{tamstorf_discretization-error-accurate_2021} shares that baseline, but derives it analytically for SPD systems from linear elliptic PDEs. Ours is measured by refinement, so it applies wherever a refined run is available.

\paragraph*{Sensitivity Analysis}
Existing approaches use automatic differentiation~\cite{menon_adapt_2019} or variance decomposition~\cite{cui_pesa_2025} to compute per-variable error sensitivities. While effective for small kernels, instruction-level tracking across chaotic integration windows incurs prohibitive overhead. The adjoint estimate (\Cref{subsec:criteria-comparison}) accumulates signed contributions, so round-off that cancels within the sum leaves it below the error actually incurred. Our framework sidesteps per-variable sensitivity, trading fine-grained optimality for scalability by validating aggregate output SNR against the discretization baseline.

\paragraph*{Probabilistic Error Analysis}
Probabilistic bounds show that rounding errors in inner products grow as $\sqrt{n}$ rather than $n$~\cite{higham_new_2019}, and Discrete Stochastic Arithmetic (DSA) estimates error propagation via randomized rounding~\cite{jezequel_cadna_2008}. However, DSA requires multiple executions or heavy instrumentation. Our framework shares the probabilistic premise but avoids repeated runs by evaluating single-execution variance directly against the discretization noise floor.

\section{Discussion and Future Work}\label{sec:discussion-and-future-work}

\begin{figure}[!tb]
    \centering
    \captionsetup[subfloat]{margin=5pt}% pad subcaptions L/R so they don't touch

    \subfloat[$q_p/\mathrm{ULP}$ matches $\mathrm{Unif}(\pm\tfrac12)$: std $0.29$, excess kurtosis $-1.2$; white, signal-independent ($\max\limits_{k\ge1}|\rho|,|\mathrm{corr}(q,u)|\le0.03$).]{%
        \includegraphics[width=0.5\columnwidth]{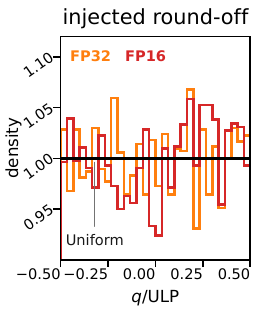}%
        \label{fig:pqn-quantizer}%
    }\hfill
    \subfloat[FP16 absorbs sub-ULP updates (${\sim}46\%$ by late time); FP32/FP64 absorb none.]{%
        \includegraphics[width=0.5\columnwidth]{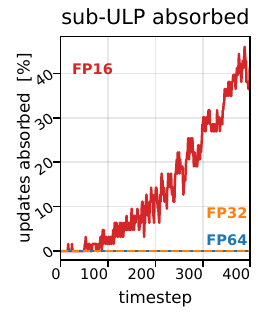}%
        \label{fig:pqn-absorption}%
    }\\[0.5ex]

    \subfloat[$t$-test of $\overline{E_{ro}}{=}0$: FP32 stays below the $5\%$ line, FP16 climbs to $t{\sim}10^2$.]{%
        \includegraphics[width=0.5\columnwidth]{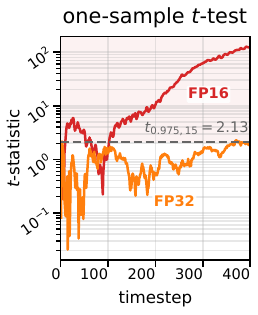}%
        \label{fig:pqn-ttest}%
    }\hfill
    \subfloat[Sign of $\overline{E_{ro}}$ across offsets: FP32 balanced ($8/16$), FP16 unanimous ($16/16$).]{%
        \includegraphics[width=0.5\columnwidth]{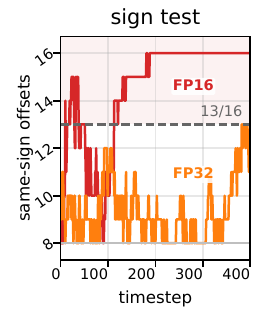}%
        \label{fig:pqn-signtest}%
    }

    \caption{Empirical test of the PQN assumptions (Burgers, FTCS). Round-off is $E_{ro}=u_p-u_{64}$, with spatial mean $\overline{E_{ro}}$. Panels~(c,~d) test whether this mean is real or a fluctuation: we re-solve at $M{=}16$ random sub-grid front offsets $\delta$ (same physics, independent round-off) and test $\overline{E_{ro}}=0$ via $t=\sqrt{M}\,\lvert\langle\overline{E_{ro}}\rangle_\delta\rvert/\sigma_\delta$ ($\langle\cdot\rangle_\delta,\sigma_\delta$: mean, stddev over offsets). PQN holds at injection~(a) but not under accumulation: the mean stays zero only while sub-ULP updates survive. FP16's signed absorption~(b) drives a real bias~(c,~d); FP32 stays unbiased far longer, borderline only at the end.}
    \label{fig:pqn}\vspace{-10pt}
\end{figure}

Our framework trades fine-grained optimality for scalability, and this choice has consequences worth discussing.
 
\paragraph*{Uniform vs.\ mixed-precision granularity}
Because we downcast uniformly, a single sensitive variable (e.g., \texttt{w\_concorr\_c} in ICON-VT) can block precision reduction for an entire subgraph. Extending the dataflow analysis to support intra-kernel precision mixing---using our coarse-grained SNR profiles to flag candidate regions for localized sensitivity analysis---would recover these missed opportunities.
 
\paragraph*{Hardware-dependent performance ceilings}
When hardware lacks native support for specific low-precision operations (e.g., FP16 transcendentals), compiler-injected format conversions erode the bandwidth benefits of a smaller footprint~\cite{heldens_kernel_2026}; FP16 VT kernel's marginal gain over FP32 despite halved memory traffic illustrates this ceiling. BF16 shows it more sharply: the same memory traffic as FP16, but division and square root lack native bfloat16 instructions and fall back to FP32 (\Cref{tab:ncu-profiling}). Emerging architectures with native BF16 datapaths shift this boundary; evaluating other representations such as block floating-point formats is a natural next step.
 
\paragraph*{Limits of the PQN assumption}
Treating round-off as uncorrelated pseudo-quantization noise~\cite{widrow_quantization_2008} is justified for chaotic, macroscopic dynamics but breaks down when errors accumulate coherently or quantization noise swamps the physical update. Figure~\ref{fig:pqn} illustrates this: the injected round-off satisfies the PQN assumptions (uniform, white, zero-mean), but once FP16's coarse ULP absorbs sub-ULP updates, the accumulated error develops a nonzero mean, while FP32 stays clean far longer. The empirical SNR checks guard against this, but only over the states and windows evaluated. Extending coverage to unseen regimes---longer horizons, broader ensembles, formal stability criteria---remains open.

\paragraph*{Beyond weather and climate}
Nothing in the criterion is specific to atmospheric models. It assumes only that the solver is discretization-limited, that a refined run can estimate the floor, and that the output is a fluctuating field whose variance is meaningful. Any stable solver run coarser than the resolution at which round-off would matter satisfies the first. It does not cover computation with no discretization to refine against; there, an arithmetic reference remains the right baseline.

\section{Conclusion}\label{sec:conclusion}

So, what \emph{does} bloat your floats? In PDE-governed simulations, over half of the 52 significand bits drown in discretization noise---no amount of arithmetic precision can recover information the numerics never had.
We formalize the intuition of ``discretization noise floor'' as a variance-based SNR, and encapsulate it in an automated DaCe-based framework that downcasts dataflow subgraphs and validates the results.
 
On ICON velocity tendencies and CLOUDSC, FP32 clears every baseline by a comfortable margin, translating to kernel speedups of up to $1.8\times$ on both platforms---close to the theoretical $2\times$ bandwidth ceiling. FP16 tells a more cautionary tale: viable in a sophisticated mixed-precision configuration (as it was for ICON-VT), and outright destructive when indiscriminately applied (as it was for CLOUDSC). BF16 does not clear the floor on either application. The safe precision boundary is application- and resolution-dependent, which is precisely why an automated criterion matters. Our framework trades per-variable optimality for scalability, the alternative being an unscalable combinatorial search. Finer-grained mixed precision, broader input coverage, and emerging block formats are natural next steps.
 
The bits discretization renders irrelevant are headroom better spent on finer grids, higher-order schemes, or larger ensembles---turning wasted arithmetic into actual science.

\section*{Acknowledgment}

This project received support from the ERC PSAP project (Grant Agreement No. 101002047). We thank the Swiss National Supercomputing Centre (CSCS) for providing computational resources and support that facilitated this project.

This work used AI assistants---Claude~\cite{anthropic_claude_2026}, Gemini~\cite{team_gemini_2025}, and Overleaf AI~\cite{overleaf_overleaf_2026}---for code development, language editing, typesetting, and bibliographic cross-referencing throughout. All outputs were verified by the authors.

\printbibliography

\end{document}